\documentstyle[side,12pt]{article}

\newcommand{\be}{\begin{equation}}
\newcommand{\ee}{\end{equation}}
\newcommand{\ben}{\begin{eqnarray}}
\newcommand{\een}{\end{eqnarray}}
\newcommand{\nn}{\nonumber}
\def\simgt{\rlap{\lower 3.5 pt\hbox{$\mathchar \sim$}}\raise 1pt \hbox {$>$}}
\def\simlt{\rlap{\lower 3.5 pt\hbox{$\mathchar \sim$}}\raise 1pt \hbox {$<$}}

\newcommand{\ssc}{\rm cont} 
\newcommand{\ssl}{\rm latt} 
\newcommand{\sla}[2]{{{#1}\hspace{-6pt}{/}}_{#2}}
\newcommand{\slasub}[2]{{{#1}\hspace{-4pt}{/}}_{#2}}
\newcommand{\slapri}[2]{{{#1}\hspace{-8pt}{/}}_{#2}}
\newcommand{\als}{\alpha_s}
\newcommand{\alsut}{\frac{\alpha_s}{u_0^3}}
\newcommand{\alsuq}{\frac{\alpha_s}{u_0^4}}
\newcommand{\intlat}{\int^{\pi}_{-\pi}\frac{d^4 k}{(2\pi)^4}}
\newcommand{\intcon}{\int^{\infty}_{-\infty}\frac{d^D k}{(2\pi)^D}}

\newcommand{\msbar}{\left(\frac{2}{\epsilon}
                    -\gamma+{\rm ln}\left|4\pi\right|\right)}
\newcommand{\logn}[1]{\hspace{3pt}{\rm ln}\left|{#1}\right|}
\newcommand{\e}[1]{$\times 10^{#1}$}

\newcommand{\ltmsq}[1]{{\Lambda^2+4{\tilde m}_{#1}^2}}
\newcommand{\tm}{\tilde m}
\newcommand{\sn}[2]{{\rm sin}({#1}_{#2})}
\newcommand{\cs}[2]{{\rm cos}({#1}_{#2})}
\newcommand{\snsq}[2]{{\rm sin}^2({#1}_{#2})}

\newcommand{\snsqh}[2]{{\rm sin}^2\left({{#1}_{#2}}/{2}\right)}

\newcommand{\sh}[2]{{\rm sinh}(E^{#1}_{#2})}
\newcommand{\ch}[2]{{\rm cosh}(E^{#1}_{#2})}

\newcommand{\shsqh}[2]{{\rm sinh}^2\left({E^{#1}_{#2}}/{2}\right)}

\newcommand{\shs}[3]{{\rm sinh}(E^{#1}_{#2}+E^{#1}_{#3})}
\newcommand{\chs}[3]{{\rm cosh}(E^{#1}_{#2}+E^{#1}_{#3})}

\newcommand{\chd}[3]{{\rm cosh}(E^{#1}_{#2}-E^{#1}_{#3})}

\begin{document}
\baselineskip=24pt
%
\title{  
{\Large Perturbative Renormalization Factors of Bilinear Operators
for Massive Wilson Quarks on the Lattice\vspace*{0.5cm}}}

\author{Yoshinobu~Kuramashi 
\vspace{5mm}\\ 
  {\it  Institute of Particle and Nuclear Studies, }\\
  {\it  High Energy Accelerator Research Organization(KEK), }\\
  {\it  Tsukuba, Ibaraki 305, Japan } 
}

\date{}
\maketitle
 
\begin{abstract}

Renormalization factors for local vector and axial vector currents 
for the Wilson quark action 
are perturbatively calculated to one loop order 
including finite quark masses
from the ratio of the on-shell quark matrix elements
in the Feynman gauge defined on the lattice and in the 
continuum. 
For large quark masses of order unity in lattice units, we find that 
finite quark mass effects are quite large: 
one-loop coefficients of the renormalization factors
differ by 100\% compared to those
in the massless limit.
  
\end{abstract}

\newpage 

\section{Introduction}
\indent

Calculation of weak matrix elements for heavy-light and
heavy-heavy mesons represents a subject of great interest 
in lattice QCD, which is in principle capable of a 
precise determination of the matrix elements
from the first principles. 
The main source of systematic errors standing in the way to 
this  goal is large $m_q a$
corrections for heavy quark mass $m_q$ in units of 
the lattice spacing $a$. In current numerical 
simulations using the Wilson quark action the magnitude of
the $b$-quark mass in lattice units is of order $m_b a\approx 1-2$.     
To control the large $m_q a$ error lattice studies have to address
two problems. 
One is improvement of the quark action to reduce cut-off effects 
following either the Symanzik approach\cite{imp_s} or Wilson's 
renormalization group approach\cite{imp_w}. 
Another problem is a precise
calculation of renormalization factors which relate
operators on the lattice to those in the continuum for massive quarks.
This calculation may be pursued either by perturbative methods\cite{rf_p} 
or by non-perturbative one\cite{rf_np}. 
An improvement of the perturbative calculation 
including the finite $m_q a$ corrections is the subject of this article.

Renormalization factors connecting the lattice 
operator to the continuum one consists of the
wave-function part and the vertex part. 
For the wave-function part 
Kronfeld and Mackenzie argued 
that the tree level normalization 
of the on-shell wave-function 
for the Wilson quark action suffers from 
large $m_q a$ corrections for 
heavy quark\cite{kmac}. 
At the one-loop level, the question of $m_qa$ corrections to the 
quark self-energy has been addressed by Kronfeld and Mertens 
for the Wilson quark action in Ref.~\cite{kmer}.
For the vertex part there are no $m_q a$ corrections
at the tree level on the lattice. 
At the one-loop level the vertex part has been calculated only in 
the massless limit so far.
For this reason analyses of weak matrix elements so far 
have to employ 
the one-loop expression in the massless limit
for the renormalization factors
combined with the Kronfeld-Mackenzie normalization 
at the tree level even for heavy quark masses.
For light quark masses $m_q a\ll 1$ we have no doubt that 
the massless expressions are a good approximation, while
for the case of heavy quark masses $m_q a\approx 1$ 
we may naturally suspect that corrections depending on $m_q a$ is 
large even at the one-loop level. 

The purpose of this paper is
to complete the one-loop calculation 
of the renormalization factors 
for the vector and axial vector currents
for finite quark masses with the Wilson quark action, and to use the 
results to investigate the magnitude of 
$m_q a$ corrections at the one-loop level. 
We calculate the renormalization factors 
of the bilinear operators  
from the ratio of the on-shell quark 
matrix elements in the Feynman gauge defined in 
the lattice regularization scheme and in the continuum 
for various combinations of the two external quark masses. 
For the continuum regularization scheme we employ
the naive dimensional regularization(NDR) with the 
${\overline {\rm MS}}$ subtraction.
In order to regularize infrared(IR) divergences
which are generated in one-loop contributions
for the on-shell wave-function renormalization factor and 
the vertex corrections,
we supply a fictitious mass $\lambda$ 
to the gluon propagator both for the lattice scheme
and the continuum one.
For the lattice scheme 
we extract the one-loop terms independent of $\lambda$
following the method used for the massless case 
in Ref.~\cite{bds}. 

This paper is organized as follows.
In Sec.~2 we give the lattice and continuum Feynman rules,
and describe the strategy for one-loop calculation 
of renormalization factors of the bilinear operators
for finite quark masses.
In Sec.~3 we demonstrate the technique to extract
the one-loop terms independent of $\lambda$
for the lattice on-shell wave-function renormalization factor, and 
present one-loop results for the relation 
between the lattice and continuum on-shell wave-function 
renormalization factors.
Results for one-loop relations
between the lattice vertex functions and 
the continuum ones are given in Sec.~4.
In Sec.~5 we present results for the renormalization factors
of the vector and axial vector currents and discuss the magnitude
of $m_q a$ corrections at the one-loop level.
In Sec.~6 we compare our results for 
the renormalization factor of the heavy-light axial vector
current with the previous static results
in the heavy quark mass region toward the static limit. 
Our conclusions are summarized in Sec.~7.

Throughout this paper we use the same notation
for quantities defined on the lattice 
and their counterparts in the continuum.  
However, in case of any possibility of confusion,
we shall make a clear distinction between them.

\section{Formalism}
\subsection{Feynman rules}
\indent

The partition function of the lattice theory
defined on a four-dimensional Euclidean space-time 
lattice with lattice spacing $a$ is given by
\ben
Z=\int \Pi_{x,\mu} {\cal D}U_\mu(x)
\Pi_{x,f}{\cal D}{\bar \psi}_f(x){\cal D}\psi_f(x)
{\rm exp}\left[-S_G-S_W\right], 
\een
where sites are labeled by $x\equiv(n_1a,n_2a,n_3a,n_4a)$
with $n_1,\cdots,n_4$ integers.
We take the standard Wilson gauge action 
for SU(3) gauge link variables $U_\mu(x)$ given by   
\ben
S_G=-\frac{1}{g^2}\sum_{x,\mu,\nu}
{\rm ReTr}\left[ U_\mu (x)U_\nu (x+{\hat \mu}) 
U_\mu^{\dagger}(x+{\hat \nu})U_\nu^{\dagger}(x)\right]
\een  
with the bare coupling constant $g$.
For the quark fields ${\bar \psi}_f(x)$ and $\psi_f(x)$ 
the Wilson quark action is given by
\ben
S_W=a^4\sum_{x,f}\left\{ \left(m_f+\frac{4r}{a}\right){\bar \psi}_f(x)
\psi_f(x)-\frac{1}{2a}\sum_\mu\left[
{\bar \psi}_f(x+{\hat \mu})(r+\gamma_\mu)
U_\mu^{\dagger}(x)\psi_f(x) 
\right.\right. \nn \\  \left.\left.
+{\bar \psi}_f(x)(r-\gamma_\mu)U_\mu(x)\psi_f(x+{\hat \mu})\right]\right\},
\een 
where ${\hat \mu}$ is a vector with length $a$ pointing
along the $\mu$-direction, 
$m_f$ is the bare quark mass for each 
flavor $f$ and $r$ is the Wilson parameter.
Color and spin indices are suppressed. We define the Euclidean 
gamma matrices in terms of the usual Minkowski
matrices in the Bjorken-Drell convention according to 
$\gamma_0=\gamma^0_{BD}$, 
$\gamma_j=-i\gamma^j_{BD}$,
$\gamma_5=\gamma^5_{BD}$; they obey 
$\{\gamma_\mu,\gamma_\nu\}=2\delta_{\mu\nu}$ and 
$\gamma_\mu^{\dagger}=\gamma_\mu$.  

Gauge link variables are elements of SU(3) group in the
fundamental representation. They can be written in the
form 
\ben 
U_\mu(x)={\rm exp}\left[ iag\sum_A T_A G^A_\mu(x)\right],
\een
where $T_A$ $(A=1,\cdots,8)$ are the generators of 
$SU(3)$ group in the fundamental representation,
which are normalized by Tr$(T_A T_B)=\delta_{AB}/2$,  
and $G^A_\mu(x)$ are the gluon fields. 
In order to derive the Feynman rules we expand  
the link variable $U_\mu(x)$ in terms of the
coupling constant $g$.
Higher order terms of $iag\sum_A T_A G^A_\mu(x)$ 
in the expansion of $U_\mu(x)$ 
yield tadpole graphs, if powers of $G^A_\mu(x)$ are contracted
with each other. These tadpole contributions are 
suppressed only by powers of $g^2$
because of cancellations between powers of $a$ from the expansion 
and ultraviolet divergences.  As a result, 
coefficients in the perturbative expansion in $g^2$
are large, and lattice perturbation theory does not converge well. 
To avoid this problem we isolate the tadpole contributions as an overall
constant $u_0$ for the expansion
of the link variable $U_\mu(x)$ in terms of $g$: 
\ben
U_\mu(x)=u_0\left[1+iag\sum_A T_A G^A_\mu(x)\right]+O(a^2),
\een 
where $u_0$ is the expectation value of the link operator in 
the gauge employed for perturbative calculations\cite{lm}. 

From now on throughout this paper
we use lattice units for expressing physical quantities
and suppress the lattice spacing $a$ unless necessary.

We summarize the lattice Feynman rules as follows.
The gluon propagator in the Feynman gauge 
with momentum $k$ is given by
\ben
\delta_{AB}\delta_{\mu\nu}u_0^4 D(k,\lambda),
\een
where 
\ben
D(k,\lambda)^{-1}=4\sum_\alpha \snsqh{k}{\alpha}+\lambda^2.
\een
Here, we give the gluon a small mass $\lambda$,  
where eventually $\lambda\rightarrow 0$,  to regularize
possible IR divergences in one-loop diagrams.
The quark propagator with momentum $k$ takes
the form
\ben
S^{-1}(k,m_u,r)=u_0\left[ i\sum_\alpha \gamma_\alpha
\sn{k}{\alpha}+m_u+2r\sum_\alpha \snsqh{k}{\alpha}\right]
\een
with $m_u=(m+4r-4ru_0)/u_0$.
The one-gluon vertex with incoming quark momentum $p$
and outgoing momentum $q$ has the following expression,
\ben
-gT_A u_0 v_\mu(p/2+q/2,r),
\een
where  
\ben
v_\mu(k,r)=i\gamma_\mu \cs{k}{\mu}+r\sn{k}{\mu}
\een
with no sum over $\mu$.
At the one-loop level the two-gluon vertex appears
only through gluon tadpole diagrams whose contributions
are included in $u_0$. 

The corresponding continuum Feynman rules 
are as follows:
the gluon propagator in the Feynman gauge is
$\delta_{AB}\delta_{\mu\nu}{\tilde D}(k,\lambda)$
with ${\tilde D}(k,\lambda)^{-1}=k^2+\lambda^2$, 
the quark propagator is given by ${\tilde S}^{-1}(k,m)=(i\sla{k}{}+m)$,
where $\sla{k}{}$ denotes $\sum_\alpha \gamma_\alpha k_\alpha$, and  
the quark-gluon vertex is $-gT_A{\tilde v}_{\mu}$ with
${\tilde v}_{\mu}=i\gamma_\mu$.

\subsection{Procedure of calculation}
\indent

In the lattice regularization scheme ultraviolet divergences 
of composite operators are regularized by the cutoff $a^{-1}$,
while in the NDR scheme in the continuum this is achieved by a 
reduction of the space-time dimension from four.
Operators defined in each regularization scheme 
can be related by renormalization factors which
are expected to converge for the perturbative expansion
in terms of $\als=g^2/(4\pi)$
due to the asymptotic freedom of the theory.
For the vector and axial vector currents the relation takes the form,
\ben
\left({\bar \psi}_2(x)\gamma_\mu\psi_1(x)\right)^{\ssc}
&=&Z_{V_\mu}\left({\bar \psi}_2(x)\gamma_\mu\psi_1(x)\right)^{\ssl},  \\
\left({\bar \psi}_2(x)\gamma_\mu\gamma_5 \psi_1(x)\right)^{\ssc}
&=&Z_{A_\mu}\left({\bar \psi}_2(x)\gamma_\mu\gamma_5 \psi_1(x)\right)^{\ssl}, 
\een
with $Z_{V_\mu}$ and $Z_{A_\mu}$ $(\mu=1,\cdots,4)$ renormalization factors. 
 
A possible way to perturbatively determine  
$Z_{V_\mu}$ and $Z_{A_\mu}$
is to calculate the ratio of the matrix elements
for the lattice and continuum regularization schemes
employing external on-shell quark or anti-quark states
in the Feynman gauge.  For the vector and axial vector 
matrix elements we further 
need to specify space momenta of external quark or anti-quark states since 
the matrix elements generally depend on them.  We take the natural choice
of zero spatial momentum in this article.  The renormalization constants 
are then calculated from
\ben
Z_{V_i}&=&\frac{\langle {\bar q}_2 
| {\bar \psi}_2(x)\gamma_i\psi_1(x) |q_1\rangle^{\ssc}}
{\langle {\bar q}_2 
| {\bar \psi}_2(x)\gamma_i\psi_1(x) |q_1\rangle^{\ssl}}
\hspace{12pt}(i=1,2,3),  
\label{eq:def_zvi} \\
Z_{V_4}&=&\frac{\langle q_2 
| {\bar \psi}_2(x)\gamma_4\psi_1(x) |q_1\rangle^{\ssc}}
{\langle q_2 
| {\bar \psi}_2(x)\gamma_4\psi_1(x) |q_1\rangle^{\ssl}},
\label{eq:def_zv4}  \\
Z_{A_i}&=&\frac{\langle q_2 
| {\bar \psi}_2(x)\gamma_i\gamma_5\psi_1(x) |q_1\rangle^{\ssc}}
{\langle q_2 
| {\bar \psi}_2(x)\gamma_i\gamma_5\psi_1(x) |q_1\rangle^{\ssl}}
\hspace{12pt}(i=1,2,3), 
\label{eq:def_zai}  \\
Z_{A_4}&=&\frac{\langle {\bar q}_2 
| {\bar \psi}_2(x)\gamma_4\gamma_5\psi_1(x) |q_1\rangle^{\ssc}}
{\langle {\bar q}_2 
| {\bar \psi}_2(x)\gamma_4\gamma_5\psi_1(x) |q_1\rangle^{\ssl}},
\label{eq:def_za4} 
\een
where $q$ and ${\bar q}$ stand for quark and anti-quark respectively.
The choice of a quark state $\langle q_2|$ or an anti-quark state 
$\langle {\bar q}_2|$ 
for the external state is made to ensure a non-zero value 
of the matrix element for each operator.  
We should note that 
because of violation of space-time permutation symmetry 
in our choice of momenta for external quark or anti-quark states
$Z_{V_i}\neq Z_{V_4}$ and $Z_{A_i}\neq Z_{A_4}$ are expected 
due to possible $m_q a$ corrections
except in the limit of $a\rightarrow 0$. 
 
At the tree level on-shell wave-function renormalization factor 
for the Wilson quark
action is shifted from unity by finite $m_q a$ corrections\cite{kmac}
\ben
&&Z_{V_i}^{(0)}=Z_{V_4}^{(0)}=Z_{A_i}^{(0)}=Z_{A_4}^{(0)}
\nn \\
&=&u_0\sqrt{\ch{(0)}{1}+r\sh{(0)}{1}}
\sqrt{\ch{(0)}{2}+r\sh{(0)}{2}},
\een
where $E^{(0)}$ denotes the pole mass at the tree level 
in common between
the Wilson quark action and the continuum one.
The superscript $(i)$ refers to the $i$-th loop level. 
Up to the one-loop level renormalization factors 
are written as
\ben 
Z_{V_i}&=&Z_{V_i}^{(0)}\left[1+\als\Delta_{V_i}\right], \\
Z_{V_4}&=&Z_{V_4}^{(0)}\left[1+\als\Delta_{V_4}\right], \\
Z_{A_i}&=&Z_{A_i}^{(0)}\left[1+\als\Delta_{A_i}\right], \\
Z_{A_4}&=&Z_{A_4}^{(0)}\left[1+\als\Delta_{A_4}\right], 
\een
with
\ben 
&&u_0=1-\frac{2}{3}g^2\intlat D(k,\lambda=0)=1-1.2976(4)\als 
\label{eq:u0}\\
&&\Delta_{V_\mu}=\frac{\Delta_{\psi_1}}{2}+\frac{\Delta_{\psi_2}}{2}
+\Delta_{\gamma_\mu}, 
\label{eq:del_vmu}\\
&&\Delta_{A_\mu}=\frac{\Delta_{\psi_1}}{2}+\frac{\Delta_{\psi_2}}{2}
+\Delta_{\gamma_\mu\gamma_5}, 
\label{eq:del_amu}
\een
where the integration in (\ref{eq:u0}) is performed
numerically using the Monte Carlo integration routine
BASES\cite{bases}. 
$\Delta_{\psi}$ is the difference at the one-loop level
between the lattice and continuum on-shell wave-function renormalization 
factors, and
$\Delta_{\Gamma}$ $(\Gamma=\gamma_\mu,\gamma_\mu\gamma_5)$ 
are a similar difference for the vertex functions.
We remark that $\Delta_{\psi}$ and $\Delta_{\Gamma}$
are functions of $E_1^{(0)}$, $E_2^{(0)}$ and $r$. 
In the following two sections we present 
our calculation of $\Delta_{\psi}$ and $\Delta_{\Gamma}$,   
and examine their pole mass dependences.

\section{Quark self-energy}
\subsection{Lattice results}
\indent

The one-loop diagram for the quark self-energy is shown
in Fig.~1,
where external quarks have zero spatial momentum.
Using the lattice Feynman 
rules in Sec.~2 we can write the one-loop contribution as
\ben
\alsut\Sigma^{(1)}(p_4,m_u,r)=\alsut\intlat I_\psi(k,p_4,m_u,r)
\een
with
\ben
I_\psi(k,p_4,m_u,r)
=4\pi C_F\sum_\rho v_\rho(p_4+k/2,r)
S(p_4+k,m_u,r) v_\rho(p_4+k/2,r)D(k,\lambda),
\label{eq:qse1}
\een
where $C_F=4/3$ is the quadratic Casimir invariant for the 
fundamental representation of SU(3) group.
The one-loop self-energy $\Sigma^{(1)}$ consists of the
kinetic and mass parts;
\ben
\Sigma^{(1)}(p_4,m_u,r)= i\gamma_4\sn{p}{4}
\Sigma^{(1)}_p(p_4,m_u,r)+\Sigma^{(1)}_m(p_4,m_u,r).
\een
With this expression the inverse of the quark propagator
up to the one-loop level is given by
\ben
S^{-1}(p_4,m_u,r)=u_0\left\{i\gamma_4\sn{p}{4}
\left[1-\alsuq\Sigma^{(1)}_p(p_4,m_u,r)\right]+m_u 
\right. \nn \\ \left.
+2r\snsqh{p}{4}-\alsuq\Sigma^{(1)}_m(p_4,m_u,r)\right\}.
\label{eq:qp_l}
\een 
Since the quark mass is additively renormalized 
at the one-loop level 
due to the chiral symmetry breaking term 
in the Wilson quark action,
on-shell condition for massless quark takes the form
$S^{-1}(p_4=0,m_u^c,r)=0$.  Here the critical quark mass $m_u^c$
measures the magnitude of the additive renormalization. It is 
determined by the integral equation
\ben
m_u^c&=&\alsuq\Sigma^{(1)}_m(p_4=0,m_u^c,r) \\
     &=&\alsuq\intlat 4\pi C_F
\frac{\left[m_u^c+2r\Delta_1\right]\Delta_6(r)+r\Delta_4}
{4\Delta_1\left\{\Delta_4+\left[m_u^c+2r\Delta_1\right]^2\right\}},
\een
where, following the notation of Ref.\cite{bds}, 
\ben
&&\Delta_1=\sum_\alpha\snsqh{k}{\alpha}, \\
&&\Delta_4=\sum_\alpha \snsq{k}{\alpha}, \\
&&\Delta_6(r)=(1+r^2)\Delta_1-4.
\een
Using $m_u^c$ we can transform (\ref{eq:qp_l}) into
the following form
\ben
S^{-1}(p_4,m_u,r)=u_0\left\{i\gamma_4\sn{p}{4}
\left[1-\alsuq\Sigma^{(1)}_p(p_4,m_u,r)\right]+m_u-m_u^c 
\right. \nn \\ \left.
+2r\snsqh{p}{4}-\alsuq\Sigma^{(1)}_m(p_4,m_u,r)
+\alsuq\Sigma^{(1)}_m(p_4=0,m_u^c,r) \right\}.
\een 
Up to the one-loop level this expression is equivalent to
\ben
S^{-1}(p_4,{\hat m}_u,r)=u_0\left\{i\gamma_4\sn{p}{4}
\left[1-\alsuq\Sigma^{(1)}_p(p_4,{\hat m}_u,r)\right]+{\hat m}_u 
\right. \nn \\ \left.
+2r\snsqh{p}{4}
-\alsuq{\hat \Sigma}^{(1)}_m(p_4,{\hat m}_u,r)\right\},
\label{eq:qp_l_mod}
\een 
where 
\ben
&&{\hat m}_u=m_u-m_u^c, \\ 
&&{\hat \Sigma}^{(1)}_m(p_4,{\hat m}_u,r)
=\Sigma^{(1)}_m(p_4,{\hat m}_u,r)-\Sigma^{(1)}_m(0,0,r),
\een
with $m_u^c=\Sigma^{(1)}_m(0,0,r)$.
We should note that, defining ${\hat m}_u$ as the bare quark mass, 
(\ref{eq:qp_l_mod}) satisfies the on-shell condition for massless
quark up to the one-loop level.
We consider that ${\hat m}_u$ corresponds
to the quark mass non-perturbatively determined  
from vanishing pion mass in Monte Carlo simulations.
In the following analysis we use ${\hat m}_u$ as
the bare quark mass.

The pole mass $E$ is given by the pole of the quark
propagator; $S^{-1}(p_4=iE,{\hat m}_u,r)=0$.
At the tree-level the equation
\ben
S^{-1}(p_4=iE^{(0)},{\hat m}_u,r)=u_0\left\{-\gamma_4\sh{(0)}{}
+{\hat m}_u-2r\shsqh{(0)}{}\right\}=0
\een 
determines
\ben
E^{(0)}=\logn{\frac{{\hat m}_u+r+\sqrt{{{\hat
m}_u}^2+2r{\hat m}_u+1}}{1+r}}.
\label{eq:m_pole}
\een
One-loop correction is obtained by
finding the pole of the quark propagator (\ref{eq:qp_l_mod}). 
Defining the one-loop term by 
\be
E=E^{(0)}+\alsuq E^{(1)}(E^{(0)},r), 
\label{eq:one-loop energy}
\ee
we find 
\ben
&&E^{(1)}(E^{(0)},r)
\nn \\
&=&\frac{\sh{(0)}{}\Sigma^{(1)}_p(p_4=iE^{(0)},{\hat m}_u,r)
-{\hat \Sigma}^{(1)}_m(p_4=iE^{(0)},{\hat m}_u,r)}
{\ch{(0)}{}+r\sh{(0)}{}}.
\label{eq:e1}
\een
For later convenience we also present the expression
of the one-loop correction to the pole mass
including the tadpole contribution, which is easily obtained 
from (\ref{eq:qp_l_mod}) with the aid of (\ref{eq:u0}),
\ben
E\,{}^{\prime}{}^{(1)}(E^{(0)},r)=E^{(1)}(E^{(0)},r)
-\frac{\sh{(0)}{}+2r\shsqh{(0)}{}}{\ch{(0)}{}+r\sh{(0)}{}}u_0^{(1)},
\label{eq:e1prime}
\een
where $u_0^{(1)}=-1.2976(4)$ from (\ref{eq:u0}).
In Table~\ref{tab:mpzwf} numerical values of 
$E^{(1)}(E^{(0)},r)$ evaluated with $r=1$ 
using BASES are given 
for representative values of $E^{(0)}$ .
The numerical accuracy is better than $2\%$.
To show the magnitude of the tadpole contribution  
we also present in Table~\ref{tab:mpzwf} 
the values of $E\,{}^{\prime}{}^{(1)}(E^{(0)},r)$
obtained from ($\ref{eq:e1prime}$).
Fig.~2 illustrate the $E^{(0)}$ dependence of $E^{(1)}$.

The on-shell wave-function renormalization factor is defined as the residue
of the quark propagator;
\ben
Z^{-1}_{\psi}(E,r)=\left. 
\frac{\partial S^{-1}(p_4,{\hat m}_u,r)}
{i\partial (p_4\gamma_4)}\right|_{p_4\gamma_4=iE}.
\een
In terms of the expression (\ref{eq:qp_l_mod}) 
$Z^{-1}_{\psi}$ is written up to the one-loop level as
\ben
Z^{-1}_{\psi}(E^{(0)},r)=u_0\left[\ch{}{}+r\sh{}{}\right]
-\alsut\left.\frac{\partial \Sigma^{(1)}(p_4,{\hat m}_u,r)}
{i\partial (p_4\gamma_4)}\right|_{p_4\gamma_4=iE^{(0)}},
\een
where $E$ is given by (\ref{eq:one-loop energy}) and 
\be
\Sigma^{(1)}(p_4,{\hat m}_u,r)= i\gamma_4\sn{p}{4}
\Sigma^{(1)}_p(p_4,{\hat m}_u,r)
+{\hat \Sigma}^{(1)}_m(p_4,{\hat m}_u,r).
\ee
Even at the tree-level the on-shell wave-function renormalization 
factor suffers from finite $m_q a$ corrections as is clear from 
\ben
{Z^{(0)}_{\psi}}^{-1}(E^{(0)},r)=u_0\left[\ch{(0)}{}+r\sh{(0)}{}\right].
\label{eq:zwf0_l}
\een
Factorizing the tree level contribution 
we obtain the following expression
\ben
\frac{Z^{-1}_{\psi}(E^{(0)},r)}{{Z^{(0)}_{\psi}}^{-1}(E^{(0)},r)}
&=&1+\alsuq\frac{\sh{(0)}{}+r\ch{(0)}{}}{\ch{(0)}{}+r\sh{(0)}{}}
E^{(1)}(E^{(0)},r)
\label{eq:zwf1_l} \\
&&-\left.\alsuq \frac{1}{\ch{(0)}{}+r\sh{(0)}{}}
\frac{\partial \Sigma^{(1)}(p_4,{\hat m}_u,r)}
{i\partial (p_4\gamma_4)}\right|_{p_4\gamma_4=iE^{(0)}}.\nn
\een

Applying the power counting rule we note 
that the right-hand side has an IR divergence
for $\lambda\rightarrow 0$
arising from the third term. 
To extract the terms independent of $\lambda$
we consider
subtracting from the integrand of the third term
an analytically integrable expression 
which has the same IR behavior in the region of small loop momentum $k$;
\ben
&&\frac{1}{\ch{(0)}{}+r\sh{(0)}{}}
\left.\frac{\partial \Sigma^{(1)}(p_4,{\hat m}_u,r)}
{i\partial (p_4\gamma_4)}\right|_{p_4\gamma_4=iE^{(0)}}
\nn \\
&=&\intlat \left. \frac{\partial {\tilde I}_{\psi}(k,p_4,E^{(0)})}
{i\partial (p_4\gamma_4)}\right|_{p_4\gamma_4=iE^{(0)}}
\label{eq:wf_l}\\
&&+\intlat \left. \frac{\partial}{i\partial (p_4\gamma_4)}
\left[\frac{I_{\psi}(k,p_4,{\hat m}_u,r)}{\ch{(0)}{}+r\sh{(0)}{}}
-{\tilde I}_{\psi}(k,p_4,E^{(0)})\right]
\right|_{p_4\gamma_4=iE^{(0)},\lambda=0},\nn
\een
where the IR behavior of
\ben
\left. \frac{\partial}{i\partial (p_4\gamma_4)} 
{\tilde I}_{\psi}(k,p_4,E^{(0)})
\right|_{p_4\gamma_4=iE^{(0)},\lambda=0}
\label{eq:iwf_c}
\een
is the same as that of
\ben  
\left. \frac{\partial}{i\partial (p_4\gamma_4)}
\frac{I_{\psi}(k,p_4,{\hat m}_u,r)}{\ch{(0)}{}+r\sh{(0)}{}}
\right|_{p_4\gamma_4=iE^{(0)},\lambda=0}.
\label{eq:iwf_l}
\een
In this expression the IR divergence is transfered
to the first term, and in consequence the second term is finite.
For the candidate of ${\tilde I}_{\psi}$ we try the continuum 
counterpart of the integrand 
$I_{\psi}/[\ch{(0)}{}+r\sh{(0)}{}]$
replacing $v_\rho(p_4+k/2,r)$ by ${\tilde v}_\rho$,
$S(p_4+k,m_u,r)$ by ${\tilde S}(p_4+k,E^{(0)})$ and $D(k,\lambda)$ 
by ${\tilde D}(k,\lambda)$ in (\ref{eq:qse1}),
\ben
{\tilde I}_\psi(k,p_4,E^{(0)})=\theta(\Lambda^2-k^2)
4\pi C_F\sum_\rho {\tilde v}_\rho {\tilde S}(p_4+k,E^{(0)})
{\tilde v}_\rho {\tilde D}(k,\lambda),
\een
where the domain of integration is restricted to a
hyper-sphere of radius $\Lambda$, not exceeding $\pi$, 
for convenience of an analytical integration. 
It is apparent that in the limit of $a\rightarrow 0$ 
(\ref{eq:iwf_c}) and (\ref{eq:iwf_l})
have the same IR behavior.
At a finite lattice spacing, however, 
the IR behaviors of the two integrands
are different due to finite $m_q a$ corrections.   
 
Let us examine the IR behaviors of the denominators
of (\ref{eq:iwf_c}) and (\ref{eq:iwf_l}),  
where we transfer the Dirac structure in the denominators of  
quark propagators to the numerators.
For the continuum case we find
\ben
&&{\tilde D}^{-1}(k,\lambda=0)\left[{\tilde S}^{-1}(p_4+k,E^{(0)}){\tilde
S}^{-1}(-(p_4+k),E^{(0)})\right]^2
\nn \\
&=&k^2\left[i2k_4 E^{(0)}+k^2\right]^2
\label{eq:den_c}
\een
with the use of the on-shell condition $p_4=iE^{(0)}$.
For the lattice case we expand
the quark and gluon propagators around $k=0$, obtaining 
\ben  
&&\left[\ch{(0)}{}+r\sh{(0)}{}\right]D^{-1}(k,\lambda=0)
\nn \\
&&\times
\left\{S^{-1}(p_4+k,{\hat m}_u,r)S^{-1}(-(p_4+k),{\hat m}_u,r)\right\}^2
\nn \\
&=&\left[\ch{(0)}{}+r\sh{(0)}{}\right]\left[k^2+O(k^4)\right]
\nn \\
&&\times\left\{i2k_4\sh{(0)}{}\left[\ch{(0)}{}+r\sh{(0)}{}\right]
+k^2(1+r\sh{(0)}{})
\right. \nn \\
&&\left.+k_4^2\sh{(0)}{}\left[(2-r^2)\sh{(0)}{}+r(\ch{(0)}{}-1)\right]
+O(k^3)\right\}^2,
\label{eq:den_l}
\een
where the on-shell condition of external quark is used.
While the two expressions above show a different IR behavior, 
the difference can be absorbed,  aside from an overall
factor, by replacing the pole mass $E^{(0)}$ in (\ref{eq:den_c}) by 
\ben
{\tilde m}={\sh{(0)}{}}\frac{\ch{(0)}{}+r\sh{(0)}{}}{1+r\sh{(0)}{}}.
\een  
Here, we do not take account of the last term in
(\ref{eq:den_l}) because it does not give 
leading order contributions either for the case of $k_4=0$
or of $k_4\neq 0$ for small $k$. 
It is straightforward to check that the remaining 
overall $m_q a$ corrections of the
denominator are precisely
canceled with those arising from the numerator with $k=0$.   
 
Consequently we take for the integrand ${\tilde I}_\psi$ 
the following expression,
\ben
{\tilde I}_\psi(k,p_4,\tm)=\theta(\Lambda^2-k^2)
4\pi C_F\sum_\rho {\tilde v}_\rho {\tilde S}(p_4+k,\tm)
{\tilde v}_\rho {\tilde D}(k,\lambda).
\een
A simple calculation gives
\ben
&&\intlat \left.\frac{\partial {\tilde I}_\psi(k,p_4,\tm)}
{i\partial (p_4\gamma_4)}\right|_{p_4\gamma_4=i\tm}
\nn \\
&=&\frac{C_F}{4\pi}\left\{
-2\logn{\frac{\lambda^2}{\Lambda^2}}
-\frac{3\Lambda^4}{4\tm^4}-\frac{9\Lambda}
{2\tm^2}\sqrt{\ltmsq{}}
+\frac{3\Lambda}{4\tm^4}(\ltmsq{})^{\frac{3}{2}}
\right.\nn \\ 
&&\left.-6\logn{\frac{\Lambda+\sqrt{\ltmsq{}}}{2\tm}}\right\},
\een
whose massless limit is
\ben
\lim_{\tm\rightarrow 0}
\intlat \left.\frac{\partial {\tilde I}_\psi(k,p_4,\tm)}
{i\partial (p_4\gamma_4)}\right|_{p_4\gamma_4=i\tm}
&=&\frac{C_F}{4\pi}\left\{
-\logn{\frac{\Lambda^2}{\tm^2}}
-2\logn{\frac{\lambda^2}{\tm^2}}
-\frac{9}{2}\right\},
\label{eq:exwf_l}
\een
where $\lambda$ is assumed to be less than ${\tilde m}$.

\subsection{Continuum results}
\indent

We turn to the calculation of the on-shell wave-function   
renormalization factor using the continuum NDR scheme instead of the lattice
regularization scheme.
For the one-loop contribution to the quark
self-energy shown in Fig.~1 the continuum 
Feynman rules in Sec.~2 give the expression 
\ben
\als \Sigma^{(1)}(p,m)=\als\intcon 
4\pi C_F\sum_\rho {\tilde v}_\rho {\tilde S}(p+k,m)
{\tilde v}_\rho {\tilde D}(k,\lambda),
\een
where $D$ is the reduced space-time dimension which
is parameterized by $\epsilon$ as
\ben
D=4-\epsilon,\hspace{12pt}\epsilon>0.
\een
Since this dimensional reduction procedure prevents us 
from taking zero spatial momentum for the
external quark state before the loop integration,
we perform the calculation of the on-shell wave-function
renormalization factor in a Euclidean invariant way.

Up to one-loop corrections 
the quark propagator takes the form
\ben
{\tilde S}^{-1}(p,m)=i\sla{p}{}+m-\als\Sigma^{(1)}(p,m).
\label{eq:qp_c}
\een
In terms of this expression 
the on-shell wave-function renormalization factor $Z^{-1}_{\psi}$ is written  
up to the one-loop level as
\ben
Z^{-1}_{\psi}(m)=1-\als\left. \frac{\partial \Sigma^{(1)}(p,m)}
{i\partial \sla{p}{}}\right|_{\slasub{p}{}=im}.
\label{eq:zwf1_c}
\een
At the tree level we find
\ben
{Z^{(0)}_\psi}^{-1}(m)=1.
\label{eq:zwf0_c}
\een
Performing the integration in an elementary way 
we obtain the one-loop correction
\ben
&&\left. \frac{\partial \Sigma^{(1)}(p,m)}
{i\partial \sla{p}{}}\right|_{\slasub{p}{}=im}
\nn \\
&=&\frac{C_F}{4\pi}\left[-\msbar
-\logn{\frac{\mu^2}{m^2}}-2\logn{\frac{\lambda^2}{m^2}}-4\right],
\label{eq:exwf_c}
\een
where the pole term 
$(2/\epsilon-\gamma+\logn{4\pi})$ should be eliminated 
in the ${\overline{\rm MS}}$ scheme.
We note that this result is independent of the choice 
of spatial momenta for the external quark state because of 
Euclidean invariance in the continuum theory.

\subsection{Relation between continuum and lattice wave-function 
renormalization factors}
\indent

Using the results for the lattice and continuum wave-function 
renormalization factors obtained in the previous subsections, 
let us find the correction factor which connects the two 
factors.
From (\ref{eq:zwf0_l}) and (\ref{eq:zwf0_c})
the tree-level relation is
\ben
{Z_{\psi}^{(0)}}^{\ssc}(m)=u_0\left[\ch{(0)}{}+r\sh{(0)}{}\right]
{Z_{\psi}^{(0)}}^{\ssl}(E^{(0)},r),
\een
where we take $m=E^{(0)}$.
Up to the one-loop level we obtain the following expression
\ben
Z_{\psi}^{\ssc}(m)=u_0\left[\ch{(0)}{}+r\sh{(0)}{}\right]
\left[1+\als\Delta_\psi(E^{(0)},r)\right]
Z_{\psi}^{\ssl}(E^{(0)},r),
\label{eq:z_wf}
\een
where, from (\ref{eq:zwf1_l}) and (\ref{eq:zwf1_c}),
\ben
\Delta_\psi(E^{(0)},r)=\left.\frac{\partial {\Sigma^{(1)}}^{\ssc}(p,m)}
{i\partial \sla{p}{}}\right|_{\slasub{p}{}=im}
+\frac{\sh{(0)}{}+r\ch{(0)}{}}{\ch{(0)}{}+r\sh{(0)}{}}
{E^{(1)}}^{\ssl}(E^{(0)},r) \nn \\ 
-\left.\frac{1}{\ch{(0)}{}+r\sh{(0)}{}}
\frac{\partial {\Sigma^{(1)}}^{\ssl}(p_4,{\hat m}_u,r)}
{i\partial (p_4\gamma_4)}\right|_{p_4\gamma_4=iE^{(0)}},
\label{eq:delta_wf}
\een
with $m=E^{(0)}$.
From (\ref{eq:e1prime}), (\ref{eq:z_wf}) and
(\ref{eq:delta_wf}) the tadpole contribution 
to the lattice wave-function
renormalization factors is explicitly expressed as
\ben
&&\Delta^{\prime}_\psi(E^{(0)},r)
\nn \\
&=&\Delta_\psi(E^{(0)},r)+u_0^{(1)}
\nn \\
&&-\frac{\left[\sh{(0)}{}+r\ch{(0)}{}\right]
\left[\sh{(0)}{}+2r\shsqh{(0)}{}\right]}
{\left[\ch{(0)}{}+r\sh{(0)}{}\right]^2}u_0^{(1)},
\label{eq:deltaprime_wf}
\een
where $u_0^{(1)}=-1.2976(4)$ from (\ref{eq:u0}).
For the coupling constant we can take either  $\als^{\ssc}$
or ${\als}^{\ssl}/u_0^4$ because the difference is of
order $\als^2$.
From (\ref{eq:exwf_l}) and (\ref{eq:exwf_c}) 
we observe that the IR singular terms
in (\ref{eq:delta_wf}) for $\lambda\rightarrow 0$ are precisely canceled. 
We also note that the mass singularities occuring at 
$E^{(0)}\rightarrow 0$ in individual terms of (\ref{eq:delta_wf}) also 
cancel, 
which assures us that $\Delta_\psi(E^{(0)},r)$ is finite even in the
massless limit. 
 
In Table~\ref{tab:mpzwf} we present numerical values of
$\Delta_\psi(E^{(0)},r)$ evaluated using BASES 
with an inaccuracy of less than $2\%$ 
for representative values of
the pole mass $E^{(0)}$ for the $r=1$ case.
The values of $\Delta^{\prime}_\psi(E^{(0)},r)$ 
defined in (\ref{eq:deltaprime_wf})
are also given in Table~\ref{tab:mpzwf} to demonstrate the magnitude 
of the tadpole contribution.  
Our result for $\Delta^\prime_\psi$ evaluated 
at $E^{(0)}=0$ with $r=1$ is 
\ben
\Delta^\prime_\psi(E^{(0)}=0,r=1)=\Delta_\psi(E^{(0)}=0,r=1)+u_0^{(1)}=-1.37,
\label{eq:delta_wf_m0}
\een
which is consistent with
$\Delta_{\Sigma_1}/(3\pi)$ for $r=1$ 
in Ref.~\cite{marti} within the error 
in the numerical integration.
The pole mass dependence of $\Delta_\psi$ 
is shown in Fig.~3. We observe that 
the $m_q a$ corrections give large contributions
in the region $E^{(0)}\simgt 0.05$.  

\section{Vertex functions}

\subsection{Lattice results}
\indent

At the tree level the lattice vertex functions for 
$\Gamma=\gamma_\mu,\gamma_\mu\gamma_5$ ($\mu=1,\cdots,4$)
do not suffer from $m_q a$ corrections, which contrasts with the
wave-function case.
Up to the one-loop level 
the vertex functions are written in the following form 
\ben
\Lambda_\Gamma(E^{(0)}_1,E^{(0)}_2,r)
=\Gamma+\alsuq\Lambda_\Gamma^{(1)}(E^{(0)}_1,E^{(0)}_2,r)
\label{eq:vf_l}
\een
where $E^{(0)}_1$ and $E^{(0)}_2$ are the pole masses 
for the external quarks.
The one-loop vertex corrections
$\Lambda_\Gamma^{(1)}$ for 
$\Gamma=\gamma_4,\gamma_i\gamma_5$ and 
$\Gamma=\gamma_i,\gamma_4\gamma_5$ ($i=1,2,3$) 
are obtained by 
evaluating the diagrams shown in Fig.~4(a)
and Fig.~4(b), respectively, 
on condition that the external quark and anti-quark
are on-shell with zero spatial momentum.

We first consider the case of 
$\Gamma=\gamma_4,\gamma_i\gamma_5$ for which external states are
quarks.
Using the lattice Feynman rules in Sec.~2, the amplitude 
corresponding to  Fig.~4(a) is 
expressed as 
\ben
\alsuq\Lambda_\Gamma^{(1)}(E^{(0)}_1,E^{(0)}_2,r)
=\alsuq\intlat I_\Gamma(k,E^{(0)}_1,E^{(0)}_2,r),
\label{eq:vc_l}
\een
with
\ben
I_\Gamma(k,E^{(0)}_1,E^{(0)}_2,r)
&=&4\pi C_F\sum_\rho v_\rho(p^{\prime}_4+k/2,r)
S(p^{\prime}_4+k,{\hat m}_{2u},r)
\nn \\
&&\times \Gamma S(p_4+k,{\hat m}_{1u},r) 
v_\rho(p_4+k/2,r)D(k,\lambda),
\label{eq:ivcv4_l}
\een
where $E^{(0)}_1$ and $E^{(0)}_2$ are 
expressed with ${\hat m}_{1u}$ and ${\hat m}_{2u}$, 
respectively, as in (\ref{eq:m_pole}).
The vertex correction (\ref{eq:vc_l}) 
has IR divergences for $\lambda\rightarrow 0$ 
as can be seen by the power counting.
Expanding (\ref{eq:ivcv4_l}) around $k=0$ we extract the IR
singular part:
\ben
&&I_{\gamma_4}(k,E^{(0)}_1,E^{(0)}_2,r)
\nn \\
&&/\left[4\pi C_F D(k,\lambda)
S(p_4+k,{\hat m}_{1u},r)S(-(p_4+k),{\hat m}_{1u},r)
\right.
\nn \\
&&\left.\times S(p^{\prime}_4+k,{\hat m}_{2u},r)
S(-(p^{\prime}_4+k),{\hat m}_{2u},r)\right]
\nn \\
&=&
\gamma_4\left[ 2\sh{(0)}{1}\sh{(0)}{2}
\left\{3-\frac{1+r^2}{2}\chs{(0)}{1}{2}
\right.\right.\nn \\
&&\left.\left.
-\frac{1-r^2}{2}\chd{(0)}{1}{2}
-r\shs{(0)}{1}{2}\right\}\right]
\nn \\
&&+\left[ 2\sh{(0)}{1}\sh{(0)}{2}
\left\{-3-\frac{1+r^2}{2}\chs{(0)}{1}{2}
\right.\right.\nn \\
&&\left.\left.
-\frac{1-r^2}{2}\chd{(0)}{1}{2}
-r\shs{(0)}{1}{2}\right\}\right]
\nn \\
&&+O(k),
\label{eq:irdiv4_l}
\\
&&I_{\gamma_i\gamma_5}(k,E^{(0)}_1,E^{(0)}_2,r)
\nn \\
&&/\left[4\pi C_F D(k,\lambda)
S(p_4+k,{\hat m}_{1u},r)S(-(p_4+k),{\hat m}_{1u},r)
\right.
\nn \\
&&\left.\times S(p^{\prime}_4+k,{\hat m}_{2u},r)
S(-(p^{\prime}_4+k),{\hat m}_{2u},r)\right]
\nn \\
&=&
\gamma_i\gamma_5\left[ 2\sh{(0)}{1}\sh{(0)}{2}
\left\{-1-\frac{1+r^2}{2}\chs{(0)}{1}{2}
\right.\right.\nn \\
&&\left.\left.
-\frac{1-r^2}{2}\chd{(0)}{1}{2}
-r\shs{(0)}{1}{2}\right\}\right]
\nn \\
&&+\gamma_i\gamma_5\gamma_4\left[ 2\sh{(0)}{1}\sh{(0)}{2}
\left\{1-\frac{1+r^2}{2}\chs{(0)}{1}{2}
\right.\right.\nn \\
&&\left.\left.
-\frac{1-r^2}{2}\chd{(0)}{1}{2}
-r\shs{(0)}{1}{2}\right\}\right]
\nn \\
&&+O(k),
\label{eq:irdivi5_l}
\een
where, from (\ref{eq:den_l}),
\ben 
&&D^{-1}(k,\lambda)
S^{-1}(p_4+k,{\hat m}_{1u},r)S^{-1}(-(p_4+k),{\hat m}_{1u},r)
\nn \\
&&\times S^{-1}(p^{\prime}_4+k,{\hat m}_{2u},r)
S^{-1}(-(p^{\prime}_4+k),{\hat m}_{2u},r)
\nn \\
&=&\left[k^2+\lambda^2+O(k^4)\right]
\nn \\
&&\times\left\{i2k_4\sh{(0)}{1}\left[\ch{(0)}{1}+r\sh{(0)}{1}\right]
\right. \nn \\
&&\left.+k^2(1+r\sh{(0)}{1})+O(k_4^2)\right\}
\nn \\
&&\times\left\{i2k_4\sh{(0)}{2}\left[\ch{(0)}{2}+r\sh{(0)}{2}\right]
\right. \nn \\
&&\left.+k^2(1+r\sh{(0)}{2})+O(k_4^2)\right\},
\een
with the use of the on-shell conditions $p_4=iE^{(0)}_1$ 
and $p_4^{\prime}=iE^{(0)}_2$.
In order to extract the terms independent of $\lambda$
we design to subtract from
the integrand $I_\Gamma(k,E^{(0)}_1,E^{(0)}_2,r)$ an
analytically integrable expression 
${\tilde I}_\Gamma$ which has the same IR behavior near
$k=0$;
\ben
\Lambda_\Gamma^{(1)}(E^{(0)}_1,E^{(0)}_2,r)
&=&\intlat {\tilde I}_{\Gamma}(k,{\tilde m}_1,{\tilde m}_2)
\label{eq:irreg} \\
&&+\intlat\left.\left[I_\Gamma(k,E^{(0)}_1,E^{(0)}_2,r)
-{\tilde I}_{\Gamma}(k,{\tilde m}_1,{\tilde m}_2)\right]
\right|_{\lambda=0}, \nn
\een
where the first term in the right hand side 
contains the IR divergence and the second one is IR finite.
For the candidate of ${\tilde I}_\Gamma$ we take
\ben
{\tilde I}_\Gamma(k,{\tilde m}_1,{\tilde m}_2)
&=&\theta(\Lambda^2-k^2)4\pi C_F  
\sum_\rho {\tilde v}_\rho {\tilde S}(p^{\prime}_4+k,{\tilde m}_2)
\nn \\
&&\times \Gamma {\tilde S}(p_4+k,{\tilde m}_1)
{\tilde v}_\rho {\tilde D}(k,\lambda),
\label{eq:vc4i5_cou}
\een
with
\ben
{\tilde m}_{1,2}=\sh{(0)}{1,2}
\frac{\ch{(0)}{1,2}+r\sh{(0)}{1,2}}{1+r\sh{(0)}{1,2}}
\label{eq:m_cou}
\een 
as we did for the case of the on-shell wave-function
renormalization factor in Sec.~3.
The IR behavior of ${\tilde I}_\Gamma$ is shown by expanding
(\ref{eq:vc4i5_cou}) around $k=0$,
\ben
&&{\tilde I}_{\gamma_4}(k,{\tilde m}_1,{\tilde m}_2)
/\left[ {4\pi C_F\tilde D}(k,\lambda)
{\tilde S}(p_4+k,{\tilde m}_1){\tilde S}(-(p_4+k),{\tilde m}_1)
\right.\nn \\
&&\left.\times{\tilde S}(p^{\prime}_4+k,{\tilde m}_2)
{\tilde S}(-(p^{\prime}_4+k),{\tilde m}_2)\right]
\nn \\
&=&\gamma_4\left[4{\tilde m}_1{\tilde m}_2\right]
+\left[-8{\tilde m}_1{\tilde m}_2\right]+O(k),
\label{eq:irdiv4_cou}
\\
&&{\tilde I}_{\gamma_i\gamma_5}(k,{\tilde m}_1,{\tilde m}_2)
/\left[ {4\pi C_F\tilde D}(k,\lambda)
{\tilde S}(p_4+k,{\tilde m}_1){\tilde S}(-(p_4+k),{\tilde m}_1)
\right.\nn \\
&&\left.\times{\tilde S}(p^{\prime}_4+k,{\tilde m}_2)
{\tilde S}(-(p^{\prime}_4+k),{\tilde m}_2)\right]
\nn \\
&=&\gamma_i\gamma_5\left[-4{\tilde m}_1{\tilde m}_2\right]+O(k),
\label{eq:irdivi5_cou}
\een
where, from (\ref{eq:den_c}),
\ben
&&{\tilde D}^{-1}(k,\lambda)
{\tilde S}^{-1}(p_4+k,{\tilde m}_1){\tilde S}^{-1}(-(p_4+k),{\tilde m}_1)
\nn \\
&&\times{\tilde S}^{-1}(p^{\prime}_4+k,{\tilde m}_2)
{\tilde S}^{-1}(-(p^{\prime}_4+k),{\tilde m}_2)
\nn \\
&=&\left[k^2+\lambda^2\right]\left[i2k_4 {\tilde m}_1+k^2\right]
\left[i2k_4 {\tilde m}_2+k^2\right],
\een
with the use of the on-shell conditions $p_4=i{\tilde m}_1$
and $p^{\prime}_4=i{\tilde m}_2$.

Comparing (\ref{eq:irdiv4_l}) with (\ref{eq:irdiv4_cou}) we find that 
each term proportional to $\gamma_4$ and $1$ in $I_{\gamma_4}$ 
shows different IR behavior from that in ${\tilde I}_{\gamma_4}$
at the finite lattice spacing $a\ne 0$.
This is not due to the operator mixing between 
the local vector current and the local scalar density but
ascribed to the lattice cut-off effects, because 
it is known that the local vector and axial vector currents 
do not mix with other dimension three 
operators for the Wilson quark action\cite{marti}. 
(In this paper we do not consider the contributions of the
higher dimensional operators to the lattice vertex functions 
(\ref{eq:vf_l}) as the operator mixing.)
Here it is noted that following our definition of the renormalization 
constant in (\ref{eq:def_zv4})
we should combine the term proportional to $\gamma_4$
with that proportional to $1$ in $I_{\gamma_4}$ and  
${\tilde I}_{\gamma_4}$ 
using the equations of motion
for the external quarks on the lattice
\ben
&& \left(i\gamma_4\sn{p}{4}+{\hat m}_{1u}+2r\snsqh{p}{4}\right)u(p_4)=0, 
\label{eq:emqq_lr}\\
&&{\bar u}(p^{\prime}_4)\left(i\gamma_4\sn{p^{\prime}}{4}
+{\hat m}_{2u}+2r\snsqh{p^{\prime}}{4}\right)=0,
\label{eq:emqq_ll}
\een
and in the continuum
\ben
&&\left(i\gamma_4 p_4+{\tilde m}_1\right)u(p_4)=0, 
\label{eq:emqq_cour}\\
&&{\bar u}(p^{\prime}_4)\left(i\gamma_4 p^{\prime}_4+{\tilde m}_2\right)=0,
\label{eq:emqq_coul}
\een
where $u(p_4)$ is the Dirac spinor 
for the incoming quark state
and  ${\bar u}(p^{\prime}_4)$ that for the outgoing quark
state.
With this procedure
we find that the IR behaviors of 
$I_{\gamma_4}$ and ${\tilde I}_{\gamma_4}$ are same, 
which is expected from a point of view that
the IR behavior should be independent of the ultra-violet 
regularization scheme.
This is the case that we can match the vertex
correction for $\Gamma=\gamma_4$ on the lattice to that in
the continuum irrespective of the IR regularization
scheme.
For $\Gamma=\gamma_i\gamma_5$ the 
combined contribution of the terms
proportional to $\gamma_i\gamma_5$ and 
$\gamma_i\gamma_5\gamma_4$ in 
$I_{\gamma_i\gamma_5}$ of (\ref{eq:irdivi5_l}) with the
aid of the
equations of motion for the external quarks 
(\ref{eq:emqq_lr}) and (\ref{eq:emqq_ll})
has the same IR behavior with 
${\tilde I}_{\gamma_i\gamma_5}$ of (\ref{eq:irdivi5_cou}).  

In the expression (\ref{eq:irreg}) 
we have carried out the first integral analytically 
and the second one numerically using BASES. 
The analytical integrations of 
${\tilde I}_\Gamma(k,{\tilde m}_1,{\tilde m}_2)$
for $\Gamma=\gamma_4,\gamma_i\gamma_5$ give the following
results
\ben
&&\intlat {\tilde I}_{\gamma_4}(k,{\tilde m}_1,{\tilde m}_2)
/\left[\frac{C_F}{4\pi}\right]
\nn \\
&=&\gamma_4\left[-2\logn{\frac{\lambda^2}{\Lambda^2}}
+\left[\left(\frac{1}{\tm_1^2}+\frac{3}{\tm_1\tm_2}\right)
\frac{\Lambda^2}{2}+\left(\frac{1}{2\tm_1^2\tm_2^2}
+\frac{1}{\tm_1^3\tm_2}\right)\frac{\Lambda^4}{4}
\right.\right.\nn \\ 
&&+\frac{\tm_2}{\tm_1-\tm_2}\left\{-4\frac{\sqrt\ltmsq{1}}{\Lambda}
+\left(\frac{1}{\tm_1^2}+\frac{2}{\tm_1\tm_2}\right)
\frac{\Lambda}{2}\sqrt\ltmsq{1}
\right. \nn \\
&&\left.\left.\left.
+\frac{\Lambda}{4\tm_1^3\tm_2}(\ltmsq{1})^{\frac{3}{2}}
+\left(6+4\frac{\tm_1}{\tm_2}\right)
\logn{\frac{\Lambda+\sqrt\ltmsq{1}}{2\tm_1}}\right\}
+\left(\tm_1 \leftrightarrow \tm_2\right)\right] \right]
\nn \\
&&+\left[4\logn{\frac{\lambda^2}{\Lambda^2}}
+\left[-\left(\frac{2}{\tm_1^2}+\frac{2}{\tm_1\tm_2}\right)
\frac{\Lambda^2}{2}
\right.\right.\nn \\ 
&&+\frac{\tm_2}{\tm_1-\tm_2}\left\{8\frac{\sqrt\ltmsq{1}}{\Lambda}
-\left(\frac{2}{\tm_1^2}+\frac{2}{\tm_1\tm_2}\right)
\frac{\Lambda}{2}\sqrt\ltmsq{1}
\right. \nn \\
&&\left.\left.\left.
-\left(12+4\frac{\tm_1}{\tm_2}\right)
\logn{\frac{\Lambda+\sqrt\ltmsq{1}}{2\tm_1}}\right\}
+\left(\tm_1 \leftrightarrow \tm_2\right)\right] \right]
\nn \\
&=&\gamma_4\left[ 2\logn{\frac{\lambda^2}{\Lambda^2}}
+\left[-\left(\frac{1}{\tm_1^2}-\frac{1}{\tm_1\tm_2}\right)
\frac{\Lambda^2}{2}+\left(\frac{1}{2\tm_1^2\tm_2^2}
+\frac{1}{\tm_1^3\tm_2}\right)\frac{\Lambda^4}{4}
\right.\right.\nn \\ 
&&-\frac{\tm_2}{\tm_1-\tm_2}\left\{-4\frac{\sqrt\ltmsq{1}}{\Lambda}
+\frac{\Lambda}{2\tm_1^2}\sqrt\ltmsq{1}
\right. \nn \\
&&\left.\left.\left.
-\frac{\Lambda}{4\tm_1^3\tm_2}(\ltmsq{1})^{\frac{3}{2}}
+6\logn{\frac{\Lambda+\sqrt\ltmsq{1}}{2\tm_1}}\right\}
+\left(\tm_1 \leftrightarrow \tm_2\right)\right] \right]
\label{eq:vc4_l}
\een
and
\ben
&&\intlat {\tilde I}_{\gamma_i\gamma_5}(k,{\tilde m}_1,{\tilde m}_2)
/\left[\frac{C_F}{4\pi}\right]
\nn \\
&=&\gamma_i\gamma_5\left[ 2\logn{\frac{\lambda^2}{\Lambda^2}}
+\left[-\left(\frac{1}{\tm_1^2}+\frac{1}{\tm_1\tm_2}\right)
\frac{\Lambda^2}{2}-\left(\frac{1}{2\tm_1^2\tm_2^2}
+\frac{1}{\tm_1^3\tm_2}\right)\frac{\Lambda^4}{12}
\right.\right. \nn \\ 
&&-\frac{\tm_2}{\tm_1-\tm_2}\left\{-4\frac{\sqrt\ltmsq{1}}{\Lambda}
+\frac{\Lambda}{2\tm_1^2}\sqrt\ltmsq{1}
\right. \nn \\
&&\left.\left.\left.
+\frac{\Lambda}{12\tm_1^3\tm_2}(\ltmsq{1})^{\frac{3}{2}}
+6\logn{\frac{\Lambda+\sqrt\ltmsq{1}}{2\tm_1}}\right\}
+\left(\tm_1 \leftrightarrow \tm_2\right)\right] \right],
\label{eq:vci5_l}
\een
where we use the equations of motion 
(\ref{eq:emqq_cour}) and (\ref{eq:emqq_coul}) 
for the second expression in (\ref{eq:vc4_l}).
We remark that above expressions do not diverge
in the limit of ${\tilde m}_2\rightarrow {\tilde m}_1$.
The results for ${\tilde m}_1={\tilde m}_2$ are presented in Appendix. 

Let us turn to the second case $\Gamma=\gamma_i,\gamma_4\gamma_5$ for 
which a quark and an anti-quark have to be chosen for the external 
states for a non-vanishing matrix element.
From Fig.~4(b) we obtain  (\ref{eq:vc_l}) with, however, 
\ben
I_\Gamma(k,E^{(0)}_1,E^{(0)}_2,r)
&=&4\pi C_F\sum_\rho v_\rho(-p^{\prime}_4+k/2,r)
S(-p^{\prime}_4+k,{\hat m}_{2u},r)
\nn \\ 
&&\times \Gamma S(p_4+k,{\hat m}_{1u},r) 
v_\rho(p_4+k/2,r)D(k,\lambda).
\label{eq:ivcvi_l}
\een
Around $k=0$ the integrand (\ref{eq:ivcvi_l}) behaves as
\ben
&&I_{\gamma_i}(k,E^{(0)}_1,E^{(0)}_2,r)
\nn \\
&&/\left[4\pi C_F D(k,\lambda)
S(p_4+k,{\hat m}_{1u},r)S(-(p_4+k),{\hat m}_{1u},r)
\right.
\nn \\
&&\left.\times S(-p^{\prime}_4+k,{\hat m}_{2u},r)
S(-(-p^{\prime}_4+k),{\hat m}_{2u},r)\right]
\nn \\
&=&
\gamma_i\left[ 2\sh{(0)}{1}\sh{(0)}{2}
\left\{1+\frac{1+r^2}{2}\chs{(0)}{1}{2}
\right.\right.\nn \\
&&\left.\left.
+\frac{1-r^2}{2}\chd{(0)}{1}{2}
+r\shs{(0)}{1}{2}\right\}\right]
\nn \\
&&+\gamma_i\gamma_4\left[ 2\sh{(0)}{1}\sh{(0)}{2}
\left\{-1+\frac{1+r^2}{2}\chs{(0)}{1}{2}
\right.\right.\nn \\
&&\left.\left.
+\frac{1-r^2}{2}\chd{(0)}{1}{2}
+r\shs{(0)}{1}{2}\right\}\right]
\nn \\
&&+O(k),
\label{eq:irdivi_l}
\\
&&I_{\gamma_4\gamma_5}(k,E^{(0)}_1,E^{(0)}_2,r)
\nn \\
&&/\left[4\pi C_F D(k,\lambda)
S(p_4+k,{\hat m}_{1u},r)S(-(p_4+k),{\hat m}_{1u},r)
\right.
\nn \\
&&\left.\times S(-p^{\prime}_4+k,{\hat m}_{2u},r)
S(-(-p^{\prime}_4+k),{\hat m}_{2u},r)\right]
\nn \\
&=&
\gamma_4\gamma_5\left[ 2\sh{(0)}{1}\sh{(0)}{2}
\left\{-3+\frac{1+r^2}{2}\chs{(0)}{1}{2}
\right.\right.\nn \\
&&\left.\left.
+\frac{1-r^2}{2}\chd{(0)}{1}{2}
+r\shs{(0)}{1}{2}\right\}\right]
\nn \\
&&+\gamma_5\left[ 2\sh{(0)}{1}\sh{(0)}{2}
\left\{-3-\frac{1+r^2}{2}\chs{(0)}{1}{2}
\right.\right.\nn \\
&&\left.\left.
-\frac{1-r^2}{2}\chd{(0)}{1}{2}
-r\shs{(0)}{1}{2}\right\}\right]
\nn \\
&&+O(k),
\label{eq:irdiv45_l}
\een
where
\ben 
&&D^{-1}(k,\lambda)
S^{-1}(p_4+k,{\hat m}_{1u},r)S^{-1}(-(p_4+k),{\hat m}_{1u},r)
\nn \\
&&\times S^{-1}(-p^{\prime}_4+k,{\hat m}_{2u},r)
S^{-1}(-(-p^{\prime}_4+k),{\hat m}_{2u},r)
\nn \\
&=&\left[k^2+\lambda^2+O(k^4)\right]
\nn \\
&&\times\left\{i2k_4\sh{(0)}{1}\left[\ch{(0)}{1}+r\sh{(0)}{1}\right]
\right. \nn \\
&&\left.+k^2(1+r\sh{(0)}{1})+O(k_4^2)\right\}
\nn \\
&&\times\left\{-i2k_4\sh{(0)}{2}\left[\ch{(0)}{2}+r\sh{(0)}{2}\right]
\right. \nn \\
&&\left.+k^2(1+r\sh{(0)}{2})+O(k_4^2)\right\},
\een
with the use of the on-shell conditions $p_4=iE^{(0)}_1$ 
and $p_4^{\prime}=iE^{(0)}_2$.

We introduce a counter term 
${\tilde I}_\Gamma(k,\tm_1,\tm_2)$ 
to regularize the
IR singularity of the integrand $I_\Gamma$ 
in the vertex correction,
\ben
{\tilde I}_\Gamma(k,{\tilde m}_1,{\tilde m}_2)
&=&\theta(\Lambda^2-k^2)4\pi C_F  
\sum_\rho {\tilde v}_\rho {\tilde S}(-p^{\prime}_4+k,{\tilde m}_2)
\nn \\
&&\times \Gamma {\tilde S}(p_4+k,{\tilde m}_1)
{\tilde v}_\rho {\tilde D}(k,\lambda)
\label{eq:vci45_cou}
\een
with $\tm_{1,2}$ defined in (\ref{eq:m_cou}).
The IR behavior of ${\tilde I}_\Gamma$ is 
\ben
&&{\tilde I}_{\gamma_i}(k,{\tilde m}_1,{\tilde m}_2)
/\left[ {4\pi C_F\tilde D}(k,\lambda)
{\tilde S}(p_4+k,{\tilde m}_1){\tilde S}(-(p_4+k),{\tilde m}_1)
\right.\nn \\
&&\left.\times{\tilde S}(-p^{\prime}_4+k,{\tilde m}_2)
{\tilde S}(-(-p^{\prime}_4+k),{\tilde m}_2)\right]
\nn \\
&=&\gamma_i\left[4{\tilde m}_1{\tilde m}_2\right]+O(k),
\label{eq:irdivi_cou}
\\
&&{\tilde I}_{\gamma_4\gamma_5}(k,{\tilde m}_1,{\tilde m}_2)
/\left[ {4\pi C_F\tilde D}(k,\lambda)
{\tilde S}(p_4+k,{\tilde m}_1){\tilde S}(-(p_4+k),{\tilde m}_1)
\right.\nn \\
&&\left.\times{\tilde S}(-p^{\prime}_4+k,{\tilde m}_2)
{\tilde S}(-(-p^{\prime}_4+k),{\tilde m}_2)\right]
\nn \\
&=&\gamma_4\gamma_5\left[-4{\tilde m}_1{\tilde m}_2\right]
+\gamma_5\left[-8{\tilde m}_1{\tilde m}_2\right]+O(k),
\label{eq:irdiv45_cou}
\een
where
\ben
&&{\tilde D}^{-1}(k,\lambda)
{\tilde S}^{-1}(p_4+k,{\tilde m}_1){\tilde S}^{-1}(-(p_4+k),{\tilde m}_1)
\nn \\
&&\times{\tilde S}^{-1}(-p^{\prime}_4+k,{\tilde m}_2)
{\tilde S}^{-1}(-(-p^{\prime}_4+k),{\tilde m}_2)
\nn \\
&=&\left[k^2+\lambda^2\right]\left[i2k_4 {\tilde m}_1+k^2\right]
\left[-i2k_4 {\tilde m}_2+k^2\right],
\een
with the use of the on-shell conditions $p_4=i{\tilde m}_1$
and $p^{\prime}_4=i{\tilde m}_2$.

As we did in the case of the vertex corrections for
$\Gamma=\gamma_4,\gamma_i\gamma_5$,
we make a one-loop matching 
of the vertex corrections for 
$\Gamma=\gamma_i,\gamma_4\gamma_5$ on the lattice to those
in the continuum with the aid of 
the equations of motion 
for the external quark and
anti-quark states on the lattice
\ben
&& \left(i\gamma_4\sn{p}{4}+{\hat m}_{1u}+2r\snsqh{p}{4}\right)u(p_4)=0, 
\label{eq:emqa_lr}\\
&&{\bar v}(p^{\prime}_4)\left(-i\gamma_4\sn{p^{\prime}}{4}
+{\hat m}_{2u}+2r\snsqh{p^{\prime}}{4}\right)=0,
\label{eq:emqa_ll}
\een
and in the continuum
\ben
&&\left(i\gamma_4 p_4+{\tilde m}_1\right)u(p_4)=0, 
\label{eq:emqa_cour}\\
&&{\bar v}(p^{\prime}_4)\left(-i\gamma_4 p^{\prime}_4+{\tilde m}_2\right)=0,
\label{eq:emqa_coul}
\een
where $u(p_4)$ is the Dirac spinor 
for the incoming quark state
and  ${\bar v}(p^{\prime}_4)$ is that 
for the incoming anti-quark state.

Using ${\tilde I}_\Gamma(k,{\tilde m}_1,{\tilde m}_2)$ 
of (\ref{eq:vci45_cou}) 
we can decompose 
the vertex correction 
$\Lambda_\Gamma^{(1)}(E^{(0)}_1,E^{(0)}_2,r)$ 
into an IR divergent part
and a finite one as in (\ref{eq:irreg}). 
The IR divergence residing in $(\ref{eq:vci45_cou})$
are calculated analytically. The finite term
is evaluated with numerical integration with the aid of BASES.
The results of an analytical integration of 
$(\ref{eq:vci45_cou})$ for 
$\Gamma=\gamma_i,\gamma_4\gamma_5$ are as follows,
\ben
&&\intlat {\tilde I}_{\gamma_i}(k,{\tilde m}_1,{\tilde m}_2)
/\left[\frac{C_F}{4\pi}\right]
\nn \\
&=&\gamma_i\left[ 2\logn{\frac{\lambda^2}{\Lambda^2}}
+8\frac{\pi}{\lambda}\frac{\tm_1\tm_2}{\tm_1+\tm_2}
\right.\nn \\
&&+\left[-\left(\frac{1}{\tm_1^2}-\frac{1}{\tm_1\tm_2}\right)
\frac{\Lambda^2}{2}-\left(\frac{1}{2\tm_1^2\tm_2^2}
-\frac{1}{\tm_1^3\tm_2}\right)\frac{\Lambda^4}{12}
\right. \nn \\ 
&&+\frac{\tm_2}{\tm_1+\tm_2}\left\{-4\frac{\sqrt\ltmsq{1}}{\Lambda}
+\frac{\Lambda}{2\tm_1^2}\sqrt\ltmsq{1}
\right. \nn \\
&&\left.\left.\left.
-\frac{\Lambda}{12\tm_1^3\tm_2}(\ltmsq{1})^{\frac{3}{2}}
+6\logn{\frac{\Lambda+\sqrt\ltmsq{1}}{2\tm_1}}\right\}
+\left( \tm_1\leftrightarrow \tm_2\right)
\right] \right]
\label{eq:vci_l}
\een
and
\ben
&&\intlat {\tilde I}_{\gamma_4\gamma_5}(k,{\tilde m}_1,{\tilde m}_2)
/\left[\frac{C_F}{4\pi}\right]
\nn \\
&=&-\gamma_4\gamma_5\left[ 2\logn{\frac{\lambda^2}{\Lambda^2}}
+8\frac{\pi}{\lambda}\frac{\tm_1\tm_2}{\tm_1+\tm_2}
\right.\nn \\
&&+\left[-\left(\frac{1}{\tm_1^2}-\frac{3}{\tm_1\tm_2}\right)
\frac{\Lambda^2}{2}+\left(-\frac{1}{2\tm_1^2\tm_2^2}
+\frac{1}{\tm_1^3\tm_2}\right)\frac{\Lambda^4}{4}
\right. \nn \\ 
&&+\frac{\tm_2}{\tm_1+\tm_2}\left\{-4\frac{\sqrt\ltmsq{1}}{\Lambda}
+\left(\frac{1}{\tm_1^2}-\frac{2}{\tm_1\tm_2}\right)
\frac{\Lambda}{2}\sqrt\ltmsq{1}
\right. \nn \\
&&\left.\left.\left.
-\frac{\Lambda}{4\tm_1^3\tm_2}(\ltmsq{1})^{\frac{3}{2}}
+\left(6-4\frac{\tm_1}{\tm_2}\right)
\logn{\frac{\Lambda+\sqrt\ltmsq{1}}{2\tm_1}}\right\}
+\left( \tm_1\leftrightarrow \tm_2\right)
\right] \right].
\nn \\
&&-\gamma_5\left[ 4\logn{\frac{\lambda^2}{\Lambda^2}}
+16\frac{\pi}{\lambda}\frac{\tm_1\tm_2}{\tm_1+\tm_2}
\right.\nn \\
&&+\left[-\left(\frac{2}{\tm_1^2}-\frac{2}{\tm_1\tm_2}\right)
\frac{\Lambda^2}{2}
\right. \nn \\ 
&&+\frac{\tm_2}{\tm_1+\tm_2}\left\{-8\frac{\sqrt\ltmsq{1}}{\Lambda}
+\left(\frac{2}{\tm_1^2}-\frac{2}{\tm_1\tm_2}\right)
\frac{\Lambda}{2}\sqrt\ltmsq{1}
\right. \nn \\
&&\left.\left.\left.
+\left(12-4\frac{\tm_1}{\tm_2}\right)
\logn{\frac{\Lambda+\sqrt\ltmsq{1}}{2\tm_1}}\right\}
+\left( \tm_1\leftrightarrow \tm_2\right)
\right] \right].
\nn \\
&=&\gamma_4\gamma_5\left[ 2\logn{\frac{\lambda^2}{\Lambda^2}}
+8\frac{\pi}{\lambda}\frac{\tm_1\tm_2}{\tm_1+\tm_2}
\right.\nn \\
&&+\left[-\left(\frac{1}{\tm_1^2}+\frac{1}{\tm_1\tm_2}\right)
\frac{\Lambda^2}{2}+\left(\frac{1}{2\tm_1^2\tm_2^2}
-\frac{1}{\tm_1^3\tm_2}\right)\frac{\Lambda^4}{4}
\right. \nn \\ 
&&+\frac{\tm_2}{\tm_1+\tm_2}\left\{-4\frac{\sqrt\ltmsq{1}}{\Lambda}
+\frac{\Lambda}{2\tm_1^2}\sqrt\ltmsq{1}
\right. \nn \\
&&\left.\left.\left.
+\frac{\Lambda}{4\tm_1^3\tm_2}(\ltmsq{1})^{\frac{3}{2}}
+6\logn{\frac{\Lambda+\sqrt\ltmsq{1}}{2\tm_1}}\right\}
+\left( \tm_1\leftrightarrow \tm_2\right)
\right] \right],
\label{eq:vc45_l}
\een
where we use the equations of motion 
(\ref{eq:emqa_cour}) and (\ref{eq:emqa_coul}) 
for the second expression in (\ref{eq:vc45_l}).
The expressions in the limit of $\tm_2\rightarrow \tm_1$ 
are listed in Appendix.

\subsection{Continuum results}
\indent

We repeat the calculations of the vertex functions for 
$\Gamma=\gamma_\mu,\gamma_\mu\gamma_5$ ($\mu=1,\cdots,4$)
in the continuum NDR scheme. 
The vertex functions up to the one-loop level 
are written as
\ben
\Lambda_\Gamma(m_1,m_2)=\Gamma+\als\Lambda_\Gamma^{(1)}(m_1,m_2).
\label{eq:vf_c}
\een
The one-loop contributions $\Lambda_\Gamma^{(1)}$ 
for $\Gamma=\gamma_4,\gamma_i\gamma_5$ and 
$\Gamma=\gamma_i,\gamma_4\gamma_5$ ($i=1,2,3$)
are represented in Fig.~4(a) and Fig.~4(b), respectively. 
We perform the loop integrations 
imposing the on-shell conditions 
on the external quark and anti-quark states
in a Euclidean invariant way, and after the integrations 
set the spatial momenta
of the external quark and anti-quark equal to zero.

For the case of $\Gamma=\gamma_4,\gamma_i\gamma_5$
the continuum Feynman rules in Sec.~2 
give the following expressions for Fig.~4(a),
\ben
\als\Lambda_\Gamma^{(1)}(m_1,m_2)
&=&\als\intcon 4\pi C_F\sum_\rho {\tilde v}_\rho
{\tilde S}(p^{\prime}+k,m_2)
\nn \\
&&\times \Gamma {\tilde S}(p+k,m_1) 
{\tilde v}_\rho{\tilde D}(k,\lambda),
\label{eq:vc_c}
\een
where $D$ is the space-time dimension which
is reduced from four by $\epsilon$ to regularize 
ultraviolet divergences and $\Gamma$ is defined as
$\gamma_\mu$ or $\gamma_\mu\gamma_5$ for $\mu=1,2,\cdots,D$.
The on-shell conditions for the external quarks are
written as
\ben
 (i\sla{p}{}+m_1)u(p)=0, 
\label{eq:emqq_cr}\\
{\bar u}(p^{\prime})(i\slapri{p^{\prime}}{}+m_2)=0,
\label{eq:emqq_cl}
\een
where $u(p)$ is the Dirac spinor for incoming quark state
and  ${\bar u}(p^{\prime})$ is one for outgoing quark state.
After carrying out the integration in (\ref{eq:vc_c})
we take $\gamma_\mu=\gamma_4$,
$\gamma_\mu\gamma_5=\gamma_i\gamma_5$,
$p=(0,0,0,im_1)$ and $p^{\prime}=(0,0,0,im_2)$. 
The final results are given by
\ben
&&\Lambda_{\gamma_4}^{(1)}(m_1,m_2)
/\left[\frac{C_F}{4\pi}\right]
\nn \\
&=&\gamma_4\left[ \msbar
+\logn{\frac{\mu^2}{m_1m_2}}-2\logn{\frac{\lambda^2}{m_1m_2}}
\right. \nn \\ 
&&
\left.
-5\frac{m_1+m_2}{m_1-m_2}\logn{\frac{m_1}{m_2}}
+6 \right]
\nn \\
&&+\left[4\logn{\frac{\lambda^2}{m_1m_2}}
+8\frac{m_1+m_2}{m_1-m_2}\logn{\frac{m_1}{m_2}}
-8 \right]
\nn \\
&=&\gamma_4\left[ \msbar
+\logn{\frac{\mu^2}{m_1m_2}}+2\logn{\frac{\lambda^2}{m_1m_2}}
\right. \nn \\ 
&&
\left.
+3\frac{m_1+m_2}{m_1-m_2}\logn{\frac{m_1}{m_2}}
-2 \right]
\label{eq:vc4_c}
\een
and
\ben
&&\Lambda_{\gamma_i\gamma_5}^{(1)}(m_1,m_2) 
/\left[\frac{C_F}{4\pi}\right]
\nn \\
&=&\gamma_i\gamma_5\left[ \msbar
+\logn{\frac{\mu^2}{m_1m_2}}+2\logn{\frac{\lambda^2}{m_1m_2}}
\right. \nn \\ 
&&
\left.
+3\frac{m_1+m_2}{m_1-m_2}\logn{\frac{m_1}{m_2}}
-4 \right],
\label{eq:vci5_c}
\een
where the pole term 
$(2/\epsilon-\gamma+\logn{4\pi})$ should be eliminated 
in the ${\overline{\rm MS}}$ scheme.
The second expression in (\ref{eq:vc4_c}) 
is obtained using the equations
of motion (\ref{eq:emqq_cr}) and (\ref{eq:emqq_cl}).
The expressions of (\ref{eq:vc4_c}) and (\ref{eq:vci5_c}) 
for the case of $m_2\rightarrow m_1$ are given in Appendix.

Another case is $\Gamma=\gamma_i,\gamma_4\gamma_5$.
The one-loop diagram shown in Fig.~4(b) 
is written as
\ben
\als\Lambda_\Gamma^{(1)}(m_1,m_2)
&=&\als\intcon 4\pi C_F\sum_\rho {\tilde v}_\rho
{\tilde S}(-p^{\prime}+k,m_2)
\nn \\
&&\times \Gamma {\tilde S}(p+k,m_1) 
{\tilde v}_\rho{\tilde D}(k,\lambda),
\een
where the external quark and anti-quark are on-shell
\ben
 (i\sla{p}{}+m_1)u(p)=0,
\label{eq:emqa_cr} \\
{\bar v}(p^{\prime})(-i\slapri{p^{\prime}}{}+m_2)=0,
\label{eq:emqa_cl} 
\een
with ${\bar v}(p^{\prime})$ the Dirac spinor 
for outgoing anti-quark state.
Performing the integration we obtain
\ben
&&\Lambda_{\gamma_i}^{(1)}(m_1,m_2)
/\left[\frac{C_F}{4\pi}\right] 
\nn \\
&=&\gamma_i\left[ \msbar
+\logn{\frac{\mu^2}{m_1m_2}}+2\logn{\frac{\lambda^2}{m_1m_2}}
+8\frac{\pi}{\lambda}\frac{m_1m_2}{m_1+m_2}
\right.\nn \\ 
&&
\left.
+3\frac{m_1-m_2}{m_1+m_2}\logn{\frac{m_1}{m_2}}
-4 \right]
\label{eq:vci_c}
\een
and
\ben
&&\Lambda_{\gamma_4\gamma_5}^{(1)}(m_1,m_2) 
/\left[\frac{C_F}{4\pi}\right] 
\nn \\
&=&\gamma_4\gamma_5\left[ \msbar
+\logn{\frac{\mu^2}{m_1m_2}}-2\logn{\frac{\lambda^2}{m_1m_2}}
-8\frac{\pi}{\lambda}\frac{m_1m_2}{m_1+m_2}
\right. \nn \\ 
&&
\left.
-5\frac{m_1-m_2}{m_1+m_2}\logn{\frac{m_1}{m_2}}
+6 \right]
\nn \\
&&+\gamma_5\left[-4\logn{\frac{\lambda^2}{m_1m_2}}
-16\frac{\pi}{\lambda}\frac{m_1m_2}{m_1+m_2}
-8\frac{m_1-m_2}{m_1+m_2}\logn{\frac{m_1}{m_2}}
+8 \right],
\nn \\
&=&\gamma_4\gamma_5\left[ \msbar
+\logn{\frac{\mu^2}{m_1m_2}}+2\logn{\frac{\lambda^2}{m_1m_2}}
+8\frac{\pi}{\lambda}\frac{m_1m_2}{m_1+m_2}
\right. \nn \\ 
&&
\left.
+3\frac{m_1-m_2}{m_1+m_2}\logn{\frac{m_1}{m_2}}
-2 \right],
\label{eq:vc45_c}
\een
where the pole term 
$(2/\epsilon-\gamma+\logn{4\pi})$ should be eliminated 
in the ${\overline{\rm MS}}$ scheme.
We use the equations
of motion (\ref{eq:emqa_cr}) and (\ref{eq:emqa_cl})
for the second expression in (\ref{eq:vc45_c}).

\subsection{Relation between continuum and lattice vertex functions}
\indent

From (\ref{eq:vf_l}) and (\ref{eq:vf_c}) we obtain the relation
between the lattice vertex corrections and the
continuum ones up to the one-loop level
\ben
{\Lambda_\Gamma}^{\ssc}(m_1,m_2)
=\left[1+\als\Delta_\Gamma(E^{(0)}_1,E^{(0)}_2,r)\right]
{\Lambda_\Gamma}^{\ssl}(E^{(0)}_1,E^{(0)}_2,r),
\een
with
\ben
\Delta_\Gamma(E^{(0)}_1,E^{(0)}_2,r)
={\Lambda_\Gamma^{(1)}}^{\ssc}(m_1,m_2)
-{\Lambda_\Gamma^{(1)}}^{\ssl}(E^{(0)}_1,E^{(0)}_2,r),
\label{eq:del_vtx}
\een
where we take $m_1=E^{(0)}_1$ and $m_2=E^{(0)}_2$.
In Appendix we list the expression for  
${\Lambda_\Gamma^{(1)}}^{\ssc}$ and 
${\Lambda_\Gamma^{(1)}}^{\ssl}$ in  the limit 
$E^{(0)}_1=E^{(0)}_2\rightarrow 0$ or $E^{(0)}_1\rightarrow 0$.
They show that ${\Lambda_\Gamma^{(1)}}^{\ssl}$ and 
${\Lambda_\Gamma^{(1)}}^{\ssc}$ have the same singular
structures for $\lambda\rightarrow 0$, 
$E^{(0)}_1=E^{(0)}_2\rightarrow 0$
or $E_1^{(0)}\rightarrow 0$ where we assume  
$\lambda < E_1^{(0)}, E_2^{(0)}$.
Thus $\Delta_\Gamma$ in (\ref{eq:del_vtx}) is IR finite, and also 
finite in the limit of $E^{(0)}_1=E^{(0)}_2\rightarrow 0$ 
or $E^{(0)}_1\rightarrow 0$. 
 
Numerical values of $\Delta_\Gamma(E^{(0)}_1,E^{(0)}_2,r)$
for $\Gamma=\gamma_\mu$  
evaluated using BASES 
with an accuracy of better than $2\%$
are tabulated in Table~\ref{tab:zv} for
representative values of the pole masses
$E_1^{(0)}$ and $E_2^{(0)}$ for the  $r=1$ case. 
Our results for $\Delta_{\gamma_i}$ and $\Delta_{\gamma_4}$
at $E_1^{(0)}=E_2^{(0)}=0$
are consistent with
$\Delta_{\gamma_\mu}/(3\pi)$ for $r=1$ 
in Ref.~\cite{marti} within the error 
in the numerical integration.
Fig.~5(a) shows the $E^{(0)}_2$ dependence of $\Delta_{\gamma_i}$
and $\Delta_{\gamma_4}$ for the case of
$E_1^{(0)}=E_2^{(0)}$, and Fig.~5(b) is the same as Fig.~5(a) 
for $E_1^{(0)}=0$.
For both cases
we observe that $\Delta_{\gamma_i}$ and $\Delta_{\gamma_4}$
are roughly consistent in the region $E^{(0)}_2\simlt 0.01$, 
while as $E^{(0)}_2$ becomes larger the absolute value
for $\Delta_{\gamma_i}$
increase and that for $\Delta_{\gamma_4}$ decrease, indicating
large $m_q a$ corrections.

Numerical data of $\Delta_\Gamma(E^{(0)}_1,E^{(0)}_2,r)$ 
for $\Gamma={\gamma_\mu\gamma_5}$ are listed in
Table~\ref{tab:za}.
The values of $\Delta_{\gamma_i\gamma_5}$ and 
$\Delta_{\gamma_4\gamma_5}$
at $E_1^{(0)}=E_2^{(0)}=0$ show an agreement 
with $\Delta_{\gamma_\mu\gamma_5}/(3\pi)$ for $r=1$ 
in Ref.~\cite{marti} within the error 
in the numerical integration.  
Our results for $\Delta_{\gamma_i\gamma_5}$ and 
$\Delta_{\gamma_4\gamma_5}$ are plotted in
Fig.~6(a) for the case of $E_1^{(0)}=E_2^{(0)}$ 
and in Fig.~6(b) for $E_1^{(0)}=0$.
We observe that $\Delta_{\gamma_i\gamma_5}$ and $\Delta_{\gamma_4\gamma_5}$
show an $E^{(0)}_2$ dependence similar to that for 
$\Delta_{\gamma_4}$ and $\Delta_{\gamma_i}$, respectively.

\section{Renormalization factors for vector and axial vector 
currents}
\indent

Since we have completed the calculations of $\Delta_\psi$ 
and $\Delta_\Gamma$ ($\Gamma=\gamma_\mu,\gamma_\mu\gamma_5$)
we are now ready to discuss the magnitude of finite $m_q a$ corrections
in the one-loop contributions of renormalization factors for vector 
and axial vector currents.
Combining the results for $\Delta_\psi$ and $\Delta_\Gamma$
we obtain values of $\Delta_{V_\mu}$ and $\Delta_{A_\mu}$ 
defined in (\ref{eq:del_vmu}) and (\ref{eq:del_amu}).

We show the results for $\Delta_{V_\mu}(E_1^{(0)},E_2^{(0)},r)$ 
in Fig.~7(a) for $E_1^{(0)}=E_2^{(0)}$ 
and in Fig.~7(b) for $E_1^{(0)}=0$ for the $r=1$ case. 
For both cases
the values for $\Delta_{V_i}$ are close to those for
$\Delta_{V_4}$ 
below $E^{(0)}_2\simlt 0.01$, beyond which 
$\Delta_{V_4}$ decreases in magnitude, whereas 
$\Delta_{V_i}$ increases.
In the heavy quark mass region $E^{(0)}_2\approx O(1-2)$,
where current $b$-quark simulations are performed,
the values of $\Delta_{V_i}$ and $\Delta_{V_4}$
deviate by about $100\%$ from those at $E^{(0)}_2=0$.

For the axial vector current
the results for $\Delta_{A_\mu}(E_1^{(0)},E_2^{(0)},r)$   
are plotted in Fig.~8(a) for $E_1^{(0)}=E_2^{(0)}$ 
and in Fig.~8(b) for $E_1^{(0)}=0$, again for the $r=1$ case.
We observe that the characteristic features 
for the $E_2^{(0)}$ dependence of $\Delta_{A_\mu}$ are similar to
those of $\Delta_{V_\mu}$, where $\Delta_{A_i}$ corresponds
to $\Delta_{V_4}$ and $\Delta_{A_4}$ to $\Delta_{V_i}$.
We note that $\Delta_{A_i}$ and
$\Delta_{A_4}$ also suffer from a $100\%$ $m_q a$ correction
in the heavy quark region $E^{(0)}_2\approx O(1-2)$.

\section{Large quark mass limit of renormalization factor
for heavy-light axial vector current}
\indent 

It is instructive to examine the quark mass dependence
of the renormalization factors in the heavy quark region 
toward the static limit comparing with the previous results
for the static effective theory\cite{static1,static2} 
and the nonrelativistic QCD(NRQCD)\cite{nrqcd1,nrqcd2}.
In the static case there exists only 
the calculation of the renormalization factor for the 
fourth-component of the static-light axial vector current 
using a static heavy quark and 
a massless Wilson quark\cite{static1,static2}. 
Thus in this section 
we consider the renormalization factor 
of the heavy-light axial vector
current. It is expected that our Wilson results 
should agree with the static ones in the large quark mass 
limit.
  
In Table~\ref{tab:heavyqm} we present numerical values 
for ${{E_2{}^\prime}^{(1)}}^{\ssl}(E_2^{(0)},r)$, 
$\Delta^\prime_\psi(E_2^{(0)},r)-2/\pi\logn{E_2^{(0)}}$ 
and $\Delta_{\gamma_4\gamma_5}(E_1^{(0)}=0,E_2^{(0)},r)$
evaluated with $r=1$ for representative values of $E_2^{(0)}$,
where the divergent part for the heavy quark mass limit
$E_2^{(0)}\longrightarrow \infty$ 
in $\Delta_\psi$ is subtracted as before.
For the values at $E_2^{(0)}=\infty$ we quote
the results in Ref.~\cite{static1}.
Since the static results for the one-loop correction to the
pole mass and that to the wave-function renormalization
factor in Ref.~\cite{static1} contain
the tadpole contributions, we present our results for 
${{E_2{}^\prime}^{(1)}}^{\ssl}$ defined in (\ref{eq:e1prime}) 
and $\Delta^\prime_\psi$ in (\ref{eq:deltaprime_wf}).

Figures~9, 10 and 11 show the $E_2^{(0)}$ dependence of 
${{E_2{}^\prime}^{(1)}}^{\ssl}$, 
$\Delta^\prime_\psi-2/\pi\logn{E_2^{(0)}}$ and $\Delta_{\gamma_4\gamma_5}$
respectively. For the comparison we also plot the results for 
the NRQCD\cite{nrqcd1}, where
we use the following correspondences between our notations and
theirs:
\ben
{{E_2{}^\prime}^{(1)}}^{\ssl}&\longleftrightarrow&
-4\pi a A^{\ssl}, \\
\Delta^\prime_\psi-2/\pi\logn{E_2^{(0)}}&\longleftrightarrow&
-\frac{4}{3\pi}-4\pi Z_{h}^{\ssl}, \\
\Delta_{\gamma_4\gamma_5}&\longleftrightarrow&
\frac{1}{3\pi}-4\pi V_{\gamma_4\gamma_5}^{\ssl}, 
\een 
with $E_2^{(0)}$ the continuum bare quark mass in common
between our Wilson results and the NRQCD ones.
We observe that our results become roughly consistent with
the static limits around $E_2^{(0)}\approx 5$, 
while the NRQCD results become closer to the static
limits rather slowly having larger values in 
magnitude than our results.
It is a reasonable feature that the NRQCD results
deviate from our Wilson ones toward the lighter quark
masses, since the NRQCD is well-defined only in the heavy
quark mass region.    
In terms of $\Delta^\prime_\psi(E_1^{(0)}=0,r)=-1.37$ 
in (\ref{eq:delta_wf_m0}), 
$\Delta^\prime_\psi(E_2^{(0)},r)-2/\pi\logn{E_2^{(0)}}$ and 
$\Delta_{\gamma_4\gamma_5}(E_1^{(0)}=0,E_2^{(0)},r)$ 
we construct  
$\Delta^\prime_{A_4}(E_1^{(0)}=0,E_2^{(0)},r)
-1/\pi\logn{E_2^{(0)}}$
following (\ref{eq:del_amu}), whose $E_2^{(0)}$ dependence is given in
Fig.~12. Both results for the Wilson quark action and the
NRQCD show a smooth approach to the static
limit, while the latter has larger absolute values than the
former over $E_2^{(0)}\simgt 1$.

\section{Conclusions and discussion}
\indent 

In this paper we have calculated 
the one-loop contributions for 
the renormalization factors of the 
vector and axial vector currents 
including finite quark mass corrections
using the Wilson quark action. 
We have demonstrated that the one-loop contributions 
suffer from a very large correction of $O(100\%)$ 
for the heavy quark masses of order unity  
in lattice units,
which corresponds to the situation of 
current $b$-quark simulations.
This fact tells us that  
the $m_q a$ corrections 
should be incorporated  even at the one-loop level for 
the renormalization factors of the weak operators 
containing the $b$-quarks in current numerical simulations
using the Wilson quark action.
We have also checked that the one-loop contributions for 
the renormalization factor of the heavy-light axial vector
current show an agreement with the static results
toward the heavy quark mass limit. 
 
In this work our investigation has been concentrated on
the Wilson quark action, which is the most naive, unimproved, 
quark action on the lattice.
We wonder how large the $m_q a$ corrections are for improved quark 
actions at the one-loop level. 
This point should
be examined through a calculation similar to ours including finite
quark mass contributions, which we leave for future investigations.

\section*{Acknowledgement}
\indent

We thank S.~Aoki for useful discussions, and 
A.~Ukawa for valuable comments and 
careful reading of the manuscript.
We also thank C.~Bernard for useful correspondence on revising
this manuscript.
This work is supported in
part by Grants-in-Aid of the Ministry of Education (No.08740221).

\section*{Appendix}
\indent

In this appendix we present expressions
of the continuum one-loop vertex corrections 
and the integrals 
of the counter terms introduced to regularize the IR
divergence of the one-loop lattice vertex corrections 
for some specified cases of the pole masses.

We first list expressions of 
the continuum vertex corrections 
$\Lambda^{(1)}_\Gamma(m_1,m_2)$  
($\Gamma=\gamma_\mu,\gamma_\mu\gamma_5$) 
for the case of $m_2\rightarrow m_1$ 
and $m_1\rightarrow 0$ with the
assumption $\lambda < m_1,m_2$. From (\ref{eq:vc4_c}),
(\ref{eq:vci5_c}), (\ref{eq:vci_c}) and (\ref{eq:vc45_c})
we find that the vertex corrections are written
in the following forms
\ben
&&\lim_{m_2\rightarrow m_1}\Lambda_{\gamma_i}^{(1)}(m_1,m_2) 
/\left[\frac{C_F}{4\pi}\right]
\nn \\
&=&\gamma_i\left[\msbar
+\logn{\frac{\mu^2}{m_1^2}}
+2\logn{\frac{\lambda^2}{m_1^2}}
+4\frac{\pi m_1}{\lambda}
-4\right],
\\
&&\lim_{m_2\rightarrow m_1}\Lambda_{\gamma_4}^{(1)}(m_1,m_2) 
/\left[\frac{C_F}{4\pi}\right]
\nn \\
&=&\gamma_4\left[\msbar
+\logn{\frac{\mu^2}{m_1^2}}
-2\logn{\frac{\lambda^2}{m_1^2}}
-4\right]
\nn \\
&&+\left[4\logn{\frac{\lambda^2}{m_1^2}}
+8\right]
\nn \\
&=&\gamma_4\left[\msbar
+\logn{\frac{\mu^2}{m_1^2}}
+2\logn{\frac{\lambda^2}{m_1^2}}
+4\right],
\label{eq:vc4_c_m1eqm2}
\\
&&\lim_{m_2\rightarrow m_1}\Lambda_{\gamma_i\gamma_5}^{(1)}(m_1,m_2) 
/\left[\frac{C_F}{4\pi}\right]
\nn \\
&=&\gamma_i\gamma_5\left[\msbar
+\logn{\frac{\mu^2}{m_1^2}}
+2\logn{\frac{\lambda^2}{m_1^2}}
+2\right],
\\
&&\lim_{m_2\rightarrow m_1}\Lambda_{\gamma_4\gamma_5}^{(1)}(m_1,m_2) 
/\left[\frac{C_F}{4\pi}\right]
\nn \\
&=&\gamma_4\gamma_5\left[\msbar
+\logn{\frac{\mu^2}{m_1^2}}
-2\logn{\frac{\lambda^2}{m_1^2}}
-4\frac{\pi m_1}{\lambda}
+6\right]
\nn \\
&&+\gamma_5\left[-4\logn{\frac{\lambda^2}{m_1^2}}
-8\frac{\pi m_1}{\lambda}
+8\right]
\nn \\
&=&\gamma_4\gamma_5\left[\msbar
+\logn{\frac{\mu^2}{m_1^2}}
+2\logn{\frac{\lambda^2}{m_1^2}}
+4\frac{\pi m_1}{\lambda}
-2\right]
\label{eq:vc45_c_m1eqm2}
\een
and
\ben
&&\lim_{m_1\rightarrow 0}\Lambda_{\gamma_i}^{(1)}(m_1,m_2) 
/\left[\frac{C_F}{4\pi}\right]
\nn \\
&=&\gamma_i\left[\msbar
+\logn{\frac{\mu^2}{m_1^2}}
+2\logn{\frac{\lambda^2}{m_1^2}}
+8\frac{\pi}{\lambda}\frac{m_1m_2}{m_1+m_2}
-4\right],
\\
&&\lim_{m_1\rightarrow 0}\Lambda_{\gamma_4}^{(1)}(m_1,m_2) 
/\left[\frac{C_F}{4\pi}\right]
\nn \\
&=&\gamma_4\left[\msbar
+\logn{\frac{\mu^2}{m_1^2}}
-2\logn{\frac{\lambda^2}{m_1^2}}
+4\logn{\frac{m_1}{m_2}}
+6\right]
\nn \\
&&+\left[4\logn{\frac{\lambda^2}{m_1^2}}
-4\logn{\frac{m_1}{m_2}}
-8\right]
\nn \\
&=&\gamma_4\left[\msbar
+\logn{\frac{\mu^2}{m_1^2}}
+2\logn{\frac{\lambda^2}{m_1^2}}
-2\right],
\label{eq:vc4_c_m10}
\\
&&\lim_{m_1\rightarrow 0}\Lambda_{\gamma_i\gamma_5}^{(1)}(m_1,m_2) 
/\left[\frac{C_F}{4\pi}\right]
\nn \\
&=&\gamma_i\gamma_5\left[\msbar
+\logn{\frac{\mu^2}{m_1^2}}
+2\logn{\frac{\lambda^2}{m_1^2}}
-4\right],
\\
&&\lim_{m_1\rightarrow 0}\Lambda_{\gamma_4\gamma_5}^{(1)}(m_1,m_2) 
/\left[\frac{C_F}{4\pi}\right]
\nn \\
&=&\gamma_4\gamma_5\left[\msbar
+\logn{\frac{\mu^2}{m_1^2}}
-2\logn{\frac{\lambda^2}{m_1^2}}
+4\logn{\frac{m_1}{m_2}}
-8\frac{\pi}{\lambda}\frac{m_1m_2}{m_1+m_2}
+6\right]
\nn \\
&&+\gamma_5\left[-4\logn{\frac{\lambda^2}{m_1^2}}
+4\logn{\frac{m_1}{m_2}}
-16\frac{\pi}{\lambda}\frac{m_1m_2}{m_1+m_2}
+8\right]
\nn \\
&=&\gamma_4\gamma_5\left[\msbar
+\logn{\frac{\mu^2}{m_1^2}}
+2\logn{\frac{\lambda^2}{m_1^2}}
+8\frac{\pi}{\lambda}\frac{m_1m_2}{m_1+m_2}
-2\right],
\label{eq:vc45_c_m10}
\een
where the second expressions in (\ref{eq:vc4_c_m1eqm2}) 
and (\ref{eq:vc4_c_m10}) are obtained using 
the equations of motion for the external quarks 
(\ref{eq:emqq_cr}) and (\ref{eq:emqq_cl}),
and those in (\ref{eq:vc45_c_m1eqm2}) and 
(\ref{eq:vc45_c_m10}) are with
the equations of motion for the external quark and
anti-quark (\ref{eq:emqa_cr}) and (\ref{eq:emqa_cl}).

We also enumerate expressions of
the integrals of the counter terms to the lattice 
vertex corrections
$\int^{\pi}_{-\pi}d^4 k/(2\pi)^4
{\tilde I}_{\Gamma}(k,{\tilde m}_1,{\tilde m}_2)$ 
($\Gamma=\gamma_\mu,\gamma_\mu\gamma_5$) for
the case of ${\tilde m}_2\rightarrow {\tilde m}_1$ and 
${\tilde m}_1\rightarrow 0$ with the
assumption $\lambda < {\tilde m}_1,{\tilde m}_2$.
From (\ref{eq:vc4_l}),
(\ref{eq:vci5_l}), (\ref{eq:vci_l}) and (\ref{eq:vc45_l})
we obtain the following expressions 
\ben
&&\lim_{\tm_2\rightarrow \tm_1}
\intlat {\tilde I}_{\gamma_i}(k,\tm_1,\tm_2)
/\left[\frac{C_F}{4\pi}\right]
\nn \\
&=&\gamma_i\left[ 2\logn{\frac{\lambda^2}{\Lambda^2}}
+6\logn{\frac{\Lambda+\sqrt\ltmsq{1}}{2\tm_1}}
+4\left[\frac{\pi\tm_1}{\lambda}
-\frac{\sqrt\ltmsq{1}}{\Lambda}\right]
\right.\nn \\
&&\left.
+\frac{1}{12}\left[\frac{\Lambda^4}{\tm_1^4}
-\frac{\Lambda}{\tm_1^4}(\ltmsq{1})^{\frac{3}{2}}\right]
+\frac{1}{2}\frac{\Lambda}{\tm_1^2}\sqrt\ltmsq{1} \right],
\\
&&\lim_{\tm_2\rightarrow \tm_1}
\intlat {\tilde I}_{\gamma_4}(k,\tm_1,\tm_2)
/\left[\frac{C_F}{4\pi}\right]
\nn \\
&=&\gamma_4\left[-2\logn{\frac{\lambda^2}{\Lambda^2}}
-2\logn{\frac{\Lambda+\sqrt\ltmsq{1}}{2\tm_1}}
\right.\nn\\
&&
\left.
+4\frac{\Lambda^2}{\tm_1^2}
+\frac{3}{4}\left[\frac{\Lambda^4}{\tm_1^4}
-\frac{\Lambda}{\tm_1^4}(\ltmsq{1})^{\frac{3}{2}}\right]
+\frac{1}{2}\frac{\Lambda}{\tm_1^2}\sqrt\ltmsq{1} \right]
\nn \\
&&+\left[4\logn{\frac{\lambda^2}{\Lambda^2}}
+8\logn{\frac{\Lambda+\sqrt\ltmsq{1}}{2\tm_1}}
-4\frac{\Lambda^2}{\tm_1^2}
+4\frac{\Lambda}{\tm_1^2}\sqrt\ltmsq{1} \right]
\nn \\
&=&\gamma_4\left[ 2\logn{\frac{\lambda^2}{\Lambda^2}}
+6\logn{\frac{\Lambda+\sqrt\ltmsq{1}}{2\tm_1}}
\right.\nn\\
&&
\left.
+\frac{3}{4}\left[\frac{\Lambda^4}{\tm_1^4}
-\frac{\Lambda}{\tm_1^4}(\ltmsq{1})^{\frac{3}{2}}\right]
+\frac{9}{2}\frac{\Lambda}{\tm_1^2}\sqrt\ltmsq{1} \right],
\label{eq:vc4_l_m1eqm2}
\\
&&\lim_{\tm_2\rightarrow \tm_1}
\intlat {\tilde I}_{\gamma_i\gamma_5}(k,\tm_1,\tm_2)
/\left[\frac{C_F}{4\pi}\right]
\nn \\
&=&\gamma_i\gamma_5\left[ 2\logn{\frac{\lambda^2}{\Lambda^2}}
+6\logn{\frac{\Lambda+\sqrt\ltmsq{1}}{2\tm_1}}
-2\frac{\Lambda^2}{\tm_1^2}
\right.\nn\\
&&
\left.
-\frac{1}{4}\left[\frac{\Lambda^4}{\tm_1^4}
-\frac{\Lambda}{\tm_1^4}(\ltmsq{1})^{\frac{3}{2}}\right]
+\frac{1}{2}\frac{\Lambda}{\tm_1^2}\sqrt\ltmsq{1} \right],
\\
&&\lim_{\tm_2\rightarrow \tm_1}
\intlat {\tilde I}_{\gamma_4\gamma_5}(k,\tm_1,\tm_2)
/\left[\frac{C_F}{4\pi}\right]
\nn \\
&=&-\gamma_4\gamma_5\left[ 2\logn{\frac{\lambda^2}{\Lambda^2}}
+2\logn{\frac{\Lambda+\sqrt\ltmsq{1}}{2\tm_1}}
+4\left[\frac{\pi\tm_1}{\lambda}
-\frac{\sqrt\ltmsq{1}}{\Lambda}\right]
\right.\nn\\
&&
\left.
+2\frac{\Lambda^2}{\tm_1^2}
+\frac{1}{4}\left[\frac{\Lambda^4}{\tm_1^4}
-\frac{\Lambda}{\tm_1^4}(\ltmsq{1})^{\frac{3}{2}}\right]
-\frac{1}{2}\frac{\Lambda}{\tm_1^2}\sqrt\ltmsq{1} \right]
\nn \\
&&-\gamma_5\left[ 4\logn{\frac{\lambda^2}{\Lambda^2}}
+8\logn{\frac{\Lambda+\sqrt\ltmsq{1}}{2\tm_1}}
+8\left[\frac{\pi\tm_1}{\lambda}
-\frac{\sqrt\ltmsq{1}}{\Lambda}\right] \right]
\nn \\
&=&\gamma_4\gamma_5\left[ 2\logn{\frac{\lambda^2}{\Lambda^2}}
+6\logn{\frac{\Lambda+\sqrt\ltmsq{1}}{2\tm_1}}
+4\left[\frac{\pi\tm_1}{\lambda}
-\frac{\sqrt\ltmsq{1}}{\Lambda}\right]
\right.\nn\\
&&
\left.
-2\frac{\Lambda^2}{\tm_1^2}
-\frac{1}{4}\left[\frac{\Lambda^4}{\tm_1^4}
-\frac{\Lambda}{\tm_1^4}(\ltmsq{1})^{\frac{3}{2}}\right]
+\frac{1}{2}\frac{\Lambda}{\tm_1^2}\sqrt\ltmsq{1} \right],
\label{eq:vc45_l_m1eqm2}
\een 
whose massless limits are 
\ben
&&\lim_{\tm_1\rightarrow 0}
\intlat {\tilde I}_{\gamma_i}(k,\tm_1,\tm_1)
/\left[\frac{C_F}{4\pi}\right]
\nn \\
&=&\gamma_i\left[\logn{\frac{\Lambda^2}{\tm_1^2}}
+2\logn{\frac{\lambda^2}{\tm_1^2}}
+4\frac{\pi\tm_1}{\lambda}
-\frac{7}{2}\right],
\\
&&\lim_{\tm_1\rightarrow 0}
\intlat {\tilde I}_{\gamma_4}(k,\tm_1,\tm_1)
/\left[\frac{C_F}{4\pi}\right]
\nn \\
&=&\gamma_4\left[\logn{\frac{\Lambda^2}{\tm_1^2}}
-2\logn{\frac{\lambda^2}{\tm_1^2}}
-\frac{7}{2}\right]
\nn \\
&&+\left[4\logn{\frac{\lambda^2}{\tm_1^2}}
+8\right]
\nn \\
&=&\gamma_4\left[\logn{\frac{\Lambda^2}{\tm_1^2}}
+2\logn{\frac{\lambda^2}{\tm_1^2}}
+\frac{9}{2}\right],
\label{eq:vc4_l_m1eqm20}
\\
&&\lim_{\tm_1\rightarrow 0}
\intlat {\tilde I}_{\gamma_i\gamma_5}(k,\tm_1,\tm_1)
/\left[\frac{C_F}{4\pi}\right]
\nn \\
&=&\gamma_i\gamma_5\left[\logn{\frac{\Lambda^2}{\tm_1^2}}
+2\logn{\frac{\lambda^2}{\tm_1^2}}
+\frac{5}{2}\right],
\\
&&\lim_{\tm_1\rightarrow 0}
\intlat {\tilde I}_{\gamma_4\gamma_5}(k,\tm_1,\tm_1)
/\left[\frac{C_F}{4\pi}\right]
\nn \\
&=&\gamma_4\gamma_5\left[\logn{\frac{\Lambda^2}{\tm_1^2}}
-2\logn{\frac{\lambda^2}{\tm_1^2}}
-4\frac{\pi\tm_1}{\lambda}
+\frac{13}{2}\right]
\nn \\
&&+\gamma_5\left[-4\logn{\frac{\lambda^2}{\tm_1^2}}
-8\frac{\pi\tm_1}{\lambda}
+8\right]
\nn \\
&=&\gamma_4\gamma_5\left[\logn{\frac{\Lambda^2}{\tm_1^2}}
+2\logn{\frac{\lambda^2}{\tm_1^2}}
+4\frac{\pi\tm_1}{\lambda}
-\frac{3}{2}\right],
\label{eq:vc45_l_m1eqm20}
\een 
and
\ben
&&\lim_{\tm_1\rightarrow 0}
\intlat {\tilde I}_{\gamma_i}(k,\tm_1,\tm_2)
/\left[\frac{C_F}{4\pi}\right]
\nn \\
&=&
\gamma_i\left[ \logn{\frac{\Lambda^2}{\tm_1^2}}
+2\logn{\frac{\lambda^2}{\tm_1^2}}-3
\right. \nn\\
&&
\left.
+8\frac{\pi}{\lambda}\frac{\tm_1\tm_2}{\tm_1+\tm_2}
+\frac{1}{2}\frac{\Lambda^2}{\tm_2^2}
+\frac{1}{12}\left[\frac{\Lambda^4}{\tm_2^4}
-\frac{\Lambda}{\tm_2^4}(\ltmsq{2})^{\frac{3}{2}}\right] \right],
\\
&&\lim_{\tm_1\rightarrow 0}
\intlat {\tilde I}_{\gamma_4}(k,\tm_1,\tm_2)
/\left[\frac{C_F}{4\pi}\right]
\nn \\
&=&\gamma_4\left[ -\logn{\frac{\Lambda^2}{\tm_1^2}}
-2\logn{\frac{\lambda^2}{\tm_1^2}}+3
+4\logn{\frac{\Lambda+\sqrt\ltmsq{2}}{2\tm_2}}
\right. \nn\\
&&
\left.
-\frac{5}{2}\frac{\Lambda^2}{\tm_2^2}
+\frac{\Lambda}{\tm_2^2}\sqrt\ltmsq{2} 
-\frac{1}{4}\left[\frac{\Lambda^4}{\tm_2^4}
-\frac{\Lambda}{\tm_2^4}(\ltmsq{2})^{\frac{3}{2}}\right] \right]
\nn \\
&&+\left[ 2\logn{\frac{\Lambda^2}{\tm_1^2}}
+4\logn{\frac{\lambda^2}{\tm_1^2}}-6
-4\logn{\frac{\Lambda+\sqrt\ltmsq{2}}{2\tm_2}}
\right. \nn\\
&&
\left.
+\frac{\Lambda^2}{\tm_2^2}
-\frac{\Lambda}{\tm_2^2}\sqrt\ltmsq{2} \right]
\nn \\
&=&\gamma_4\left[ \logn{\frac{\Lambda^2}{\tm_1^2}}
+2\logn{\frac{\lambda^2}{\tm_1^2}}-3
\right. \nn\\
&&
\left.
-\frac{3}{2}\frac{\Lambda^2}{\tm_2^2}
-\frac{1}{4}\left[\frac{\Lambda^4}{\tm_2^4}
-\frac{\Lambda}{\tm_2^4}(\ltmsq{2})^{\frac{3}{2}}\right] \right],
\label{eq:vc4_l_m10}
\\
&&\lim_{\tm_1\rightarrow 0}
\intlat {\tilde I}_{\gamma_i\gamma_5}(k,\tm_1,\tm_2)
/\left[\frac{C_F}{4\pi}\right]
\nn \\
&=&
\gamma_i\gamma_5\left[ \logn{\frac{\Lambda^2}{\tm_1^2}}
+2\logn{\frac{\lambda^2}{\tm_1^2}}-3
\right. \nn\\
&&
\left.
+\frac{1}{2}\frac{\Lambda^2}{\tm_2^2}
+\frac{1}{12}\left[\frac{\Lambda^4}{\tm_2^4}
-\frac{\Lambda}{\tm_2^4}(\ltmsq{2})^{\frac{3}{2}}\right] \right],
\\
&&\lim_{\tm_1\rightarrow 0}
\intlat {\tilde I}_{\gamma_4\gamma_5}(k,\tm_1,\tm_2)
/\left[\frac{C_F}{4\pi}\right]
\nn \\
&=&
\gamma_4\gamma_5\left[ -\logn{\frac{\Lambda^2}{\tm_1^2}}
-2\logn{\frac{\lambda^2}{\tm_1^2}}+3
+4\logn{\frac{\Lambda+\sqrt\ltmsq{2}}{2\tm_2}}
\right. \nn\\
&&
\left.
-8\frac{\pi}{\lambda}\frac{\tm_1\tm_2}{\tm_1+\tm_2}
-\frac{5}{2}\frac{\Lambda^2}{\tm_2^2}
+\frac{\Lambda}{\tm_2^2}\sqrt\ltmsq{2} 
-\frac{1}{4}\left[\frac{\Lambda^4}{\tm_2^4}
-\frac{\Lambda}{\tm_2^4}(\ltmsq{2})^{\frac{3}{2}}\right] \right]
\nn \\
&&+\gamma_5\left[ -2\logn{\frac{\Lambda^2}{\tm_1^2}}
-4\logn{\frac{\lambda^2}{\tm_1^2}}+6
+4\logn{\frac{\Lambda+\sqrt\ltmsq{2}}{2\tm_2}}
\right. \nn\\
&&
\left.
-16\frac{\pi}{\lambda}\frac{\tm_1\tm_2}{\tm_1+\tm_2}
-\frac{\Lambda^2}{\tm_2^2}
+\frac{\Lambda}{\tm_2^2}\sqrt\ltmsq{2} \right]
\nn \\
&=&
\gamma_4\gamma_5\left[ \logn{\frac{\Lambda^2}{\tm_1^2}}
+2\logn{\frac{\lambda^2}{\tm_1^2}}-3
\right. \nn\\
&&
\left.
+8\frac{\pi}{\lambda}\frac{\tm_1\tm_2}{\tm_1+\tm_2}
-\frac{3}{2}\frac{\Lambda^2}{\tm_2^2}
-\frac{1}{4}\left[\frac{\Lambda^4}{\tm_2^4}
-\frac{\Lambda}{\tm_2^4}(\ltmsq{2})^{\frac{3}{2}}\right] \right],
\label{eq:vc45_l_m10}
\een 
where the second expressions in 
(\ref{eq:vc4_l_m1eqm2}), (\ref{eq:vc4_l_m1eqm20}) 
and (\ref{eq:vc4_l_m10}) are obtained using 
the equations of motion for the external quarks 
(\ref{eq:emqq_cour}) and (\ref{eq:emqq_coul}),
and those in (\ref{eq:vc45_l_m1eqm2}), (\ref{eq:vc45_l_m1eqm20}) 
and (\ref{eq:vc45_l_m10}) are with
the equations of motion for the external quark and
anti-quark (\ref{eq:emqa_cour}) and (\ref{eq:emqa_coul}).

\newpage
\begin{center}
\section*{Tables}
\end{center}

\begin{table}[h]
\vspace{-1mm}
\begin{center}
\caption{\label{tab:mpzwf}$E^{(0)}$ dependence of
${E^{(1)}}^{\ssl}(E^{(0)},r)$,
${E\,{}^{\prime}{}^{(1)}}^{\ssl}(E^{(0)},r)$, 
$\Delta_\psi(E^{(0)},r)$ and $\Delta^{\prime}_\psi(E^{(0)},r)$.
The Wilson parameter $r$ is chosen to be one.
For ${E^{(1)}}^{\ssl}$ and $\Delta_\psi$ 
the error in the numerical integration is less than $2\%$.
The values for ${E\,{}^{\prime}{}^{(1)}}^{\ssl}$ and 
$\Delta^{\prime}_\psi$ are evaluated 
using (\protect{\ref{eq:e1prime}}) 
and (\protect{\ref{eq:deltaprime_wf}}).}
\vspace*{2mm}
\begin{tabular}{lllll}\hline
    $E^{(0)}$     & ${E^{(1)}}^{\ssl}$ 
                  & ${E\,{}^{\prime}{}^{(1)}}^{\ssl}$
                  & $\Delta_\psi$ 
                  & $\Delta^{\prime}_\psi$ \\ 
\hline
0       & 0          & 0            & $-$6.90\e{-2} & $-$1.37  \\
0.001   & 4.96\e{-3} & 6.26\e{-3}   & $-$6.83\e{-2} & $-$1.37  \\
0.01    & 3.39\e{-2} & 4.68\e{-2}   & $-$8.30\e{-2} & $-$1.37  \\ 
0.1     & 1.80\e{-1} & 3.04\e{-1}   & $-$1.59\e{-1} & $-$1.33  \\ 
0.2     & 2.66\e{-1} & 5.01\e{-1}   & $-$2.07\e{-1} & $-$1.27  \\ 
0.5     & 3.99\e{-1} & 9.10\e{-1}   & $-$2.83\e{-1} & $-$1.07  \\ 
1       & 5.13\e{-1} & 1.33         & $-$2.98\e{-1} & $-$7.75\e{-1}  \\ 
2       & 6.68\e{-1} & 1.79         & $-$2.43\e{-1} & $-$4.19\e{-1}  \\         
\hline
\end{tabular} 
\end{center}
\end{table}

\newpage

\begin{sidetable}[h]
\vspace{-1mm}
\begin{center}
\caption{\label{tab:zv}
(a) $\Delta_{\gamma_i}(E_1^{(0)},E_2^{(0)},r)$ and
(b) $\Delta_{\gamma_4}(E_1^{(0)},E_2^{(0)},r)$ 
for combinations of $E_1^{(0)}$ and $E_2^{(0)}$.
The Wilson parameter $r$ is chosen to be one.
The error in the numerical integration is less than $2\%$.}
\vspace*{2mm}
\begin{tabular}{lllllllll}\hline
&&&&\multicolumn{2}{l}{(a) $\Delta_{\gamma_i} (i=1,2,3)$}&& \\
&&&&\multicolumn{2}{c}{$E_1^{(0)}$}&& \\
$E_2^{(0)}$ & 0 & 0.001 & 0.01 & 0.1 & 0.2 & 0.5 & 1 & 2 \\ 
\hline
0       & $-$8.29\e{-1} &&&&&&& \\
0.001   & $-$8.30\e{-1} & $-$8.31\e{-1} &&&&&& \\
0.01    & $-$8.43\e{-1} & $-$8.44\e{-1} 
        & $-$8.58\e{-1} &&&&& \\
0.1     & $-$9.18\e{-1} & $-$9.11\e{-1} 
        & $-$9.28\e{-1} & $-$9.94\e{-1} &&&& \\
0.2     & $-$9.71\e{-1} & $-$9.66\e{-1} 
        & $-$9.82\e{-1} & $-$1.04 
        & $-$1.08       &&& \\ 
0.5     & $-$1.08       & $-$1.07 
        & $-$1.09       & $-$1.14 
        & $-$1.16       & $-$1.19       && \\ 
1       & $-$1.18       & $-$1.18 
        & $-$1.20       & $-$1.27 
        & $-$1.31       & $-$1.31 
        & $-$1.31       & \\ 
2       & $-$1.32       & $-$1.32
        & $-$1.35       & $-$1.46 
        & $-$1.54       & $-$1.69
        & $-$1.77       & $-$2.12 \\         
\multicolumn{9}{c}{} \\
&&&&\multicolumn{2}{l}{(b) $\Delta_{\gamma_4}$}&& \\
&&&&\multicolumn{2}{c}{$E_1^{(0)}$}&& \\
$E_2^{(0)}$ & 0 & 0.001 & 0.01 & 0.1 & 0.2 & 0.5 & 1 & 2 \\ 
\hline
0       & $-$8.19\e{-1} &&&&&&& \\
0.001   & $-$8.24\e{-1} & $-$8.19\e{-1} &&&&&& \\
0.01    & $-$8.13\e{-1} & $-$8.11\e{-1} 
        & $-$8.06\e{-1} &&&&& \\
0.1     & $-$7.31\e{-1} & $-$7.33\e{-1} 
        & $-$7.30\e{-1} & $-$6.92\e{-1} &&&& \\
0.2     & $-$6.50\e{-1} & $-$6.55\e{-1} 
        & $-$6.53\e{-1} & $-$6.38\e{-1} 
        & $-$6.03\e{-1} &&& \\ 
0.5     & $-$4.57\e{-1} & $-$4.51\e{-1} 
        & $-$4.62\e{-1} & $-$4.81\e{-1} 
        & $-$4.74\e{-1} & $-$4.00\e{-1} && \\ 
1       & $-$2.19\e{-1} & $-$2.16\e{-1} 
        & $-$2.18\e{-1} & $-$2.55\e{-1} 
        & $-$2.69\e{-1} & $-$2.48\e{-1} 
        & $-$1.75\e{-1} & \\ 
2       & $+$7.40\e{-2} & $+$7.20\e{-2} 
        & $+$6.96\e{-2} & $+$3.58\e{-2} 
        & $+$9.03\e{-3} & $-$2.73\e{-2} 
        & $-$2.31\e{-2} & $+$3.20\e{-2} \\         
\hline
\end{tabular} 
\end{center}
\vspace{-12mm}
\end{sidetable}

\begin{sidetable}[h]
\vspace{-1mm}
\begin{center}
\caption{\label{tab:za}
(a) $\Delta_{\gamma_i\gamma_5}(E_1^{(0)},E_2^{(0)},r)$ and
(b) $\Delta_{\gamma_4\gamma_5}(E_1^{(0)},E_2^{(0)},r)$ 
for combinations of $E_1^{(0)}$ and $E_2^{(0)}$.
The Wilson parameter $r$ is chosen to be one.
The error in the numerical integration is less than $2\%$.}
\vspace*{2mm}
\begin{tabular}{lllllllll}\hline
&&&&\multicolumn{2}{l}{(a) $\Delta_{\gamma_i\gamma_5} (i=1,2,3)$}&& \\
&&&&\multicolumn{2}{c}{$E_1^{(0)}$}&& \\
$E_2^{(0)}$ & 0 & 0.001 & 0.01 & 0.1 & 0.2 & 0.5 & 1 & 2 \\ 
\hline
0       & $-$3.13\e{-1} &&&&&&& \\
0.001   & $-$3.14\e{-1} & $-$3.12\e{-1} &&&&&& \\
0.01    & $-$3.08\e{-1} & $-$3.05\e{-1} 
        & $-$2.99\e{-1} &&&&& \\
0.1     & $-$2.71\e{-1} & $-$2.62\e{-1} 
        & $-$2.66\e{-1} & $-$2.44\e{-1} &&&& \\
0.2     & $-$2.43\e{-1} & $-$2.33\e{-1} 
        & $-$2.40\e{-1} & $-$2.27\e{-1} 
        & $-$2.17\e{-1} &&& \\ 
0.5     & $-$1.91\e{-1} & $-$1.85\e{-1} 
        & $-$1.83\e{-1} & $-$1.88\e{-1} 
        & $-$1.90\e{-1} & $-$1.88\e{-1} && \\ 
1       & $-$1.31\e{-1} & $-$1.25\e{-1} 
        & $-$1.24\e{-1} & $-$1.31\e{-1} 
        & $-$1.43\e{-1} & $-$1.58\e{-1} 
        & $-$1.49\e{-1} & \\ 
2       & $-$2.87\e{-2} & $-$3.00\e{-2} 
        & $-$3.04\e{-2} & $-$4.94\e{-2}
        & $-$6.66\e{-2} & $-$1.04\e{-1}
        & $-$1.24\e{-1} & $-$1.21\e{-1} \\         
\multicolumn{9}{c}{} \\
&&&&\multicolumn{2}{l}{(b) $\Delta_{\gamma_4\gamma_5}$}&& \\
&&&&\multicolumn{2}{c}{$E_1^{(0)}$}&& \\
$E_2^{(0)}$ & 0 & 0.001 & 0.01 & 0.1 & 0.2 & 0.5 & 1 & 2 \\ 
\hline
0       & $-$3.15\e{-1} &&&&&&& \\
0.001   & $-$3.15\e{-1} & $-$3.14\e{-1} &&&&&& \\
0.01    & $-$3.09\e{-1} & $-$3.06\e{-1} 
        & $-$3.03\e{-1} &&&&& \\
0.1     & $-$2.90\e{-1} & $-$2.83\e{-1} 
        & $-$2.82\e{-1} & $-$2.61\e{-1} &&&& \\
0.2     & $-$2.88\e{-1} & $-$2.78\e{-1} 
        & $-$2.82\e{-1} & $-$2.60\e{-1} 
        & $-$2.53\e{-1} &&& \\ 
0.5     & $-$3.36\e{-1} & $-$3.26\e{-1} 
        & $-$3.36\e{-1} & $-$3.25\e{-1} 
        & $-$3.10\e{-1} & $-$3.13\e{-1} && \\ 
1       & $-$4.83\e{-1} & $-$4.74\e{-1} 
        & $-$4.89\e{-1} & $-$5.15\e{-1} 
        & $-$5.21\e{-1} & $-$4.99\e{-1} 
        & $-$5.46\e{-1} & \\ 
2       & $-$8.16\e{-1} & $-$8.20\e{-1} 
        & $-$8.38\e{-1} & $-$9.27\e{-1} 
        & $-$1.00       & $-$1.12 
        & $-$1.20       & $-$1.61 \\         
\hline
\end{tabular} 
\end{center}
\vspace{-12mm}
\end{sidetable}

\clearpage

\begin{table}[h]
\vspace{-1mm}
\begin{center}
\caption{\label{tab:heavyqm}$E_2^{(0)}$ dependence of
${{E_2{}^\prime}^{(1)}}^{\ssl}(E_2^{(0)},r)$, 
$\Delta^\prime_\psi(E_2^{(0)},r)-2/\pi\logn{E_2^{(0)}}$ and 
$\Delta_{\gamma_4\gamma_5}(E_1^{(0)}=0,E_2^{(0)},r)$.
The Wilson parameter $r$ is chosen to be one.
The error in the numerical integration is less than $2\%$.}
\vspace*{2mm}
\begin{tabular}{llll}\hline
    $E_2^{(0)}$     & ${{E_2{}^\prime}^{(1)}}^{\ssl}$ 
& $\Delta^\prime_\psi-2/\pi\logn{E_2^{(0)}}$ 
& $\Delta_{\gamma_4\gamma_5}$\\ 
\hline
1.0    & 1.33   & $-$7.75\e{-1} & $-$4.83\e{-1} \\
1.4    & 1.53   & $-$8.26\e{-1} & $-$6.21\e{-1} \\
1.8    & 1.70   & $-$8.52\e{-1} & $-$7.55\e{-1} \\ 
2.2    & 1.81   & $-$8.81\e{-1} & $-$8.74\e{-1} \\ 
2.6    & 1.90   & $-$8.76\e{-1} & $-$9.78\e{-1} \\ 
3.0    & 1.96   & $-$8.82\e{-1} & $-$1.06       \\ 
4.0    & 2.07   & $-$8.85\e{-1} & $-$1.17       \\ 
5.0    & 2.13   & $-$9.12\e{-1} & $-$1.20       \\         
$\infty$\protect{\cite{static1}} & 2.117  & $-$9.05\e{-1} & $-$1.239  \\         
\hline
\end{tabular} 
\end{center}
\end{table}

\clearpage

\newpage
\begin{center}
\section*{Figure Captions}
\end{center}
 
\begin{itemize}
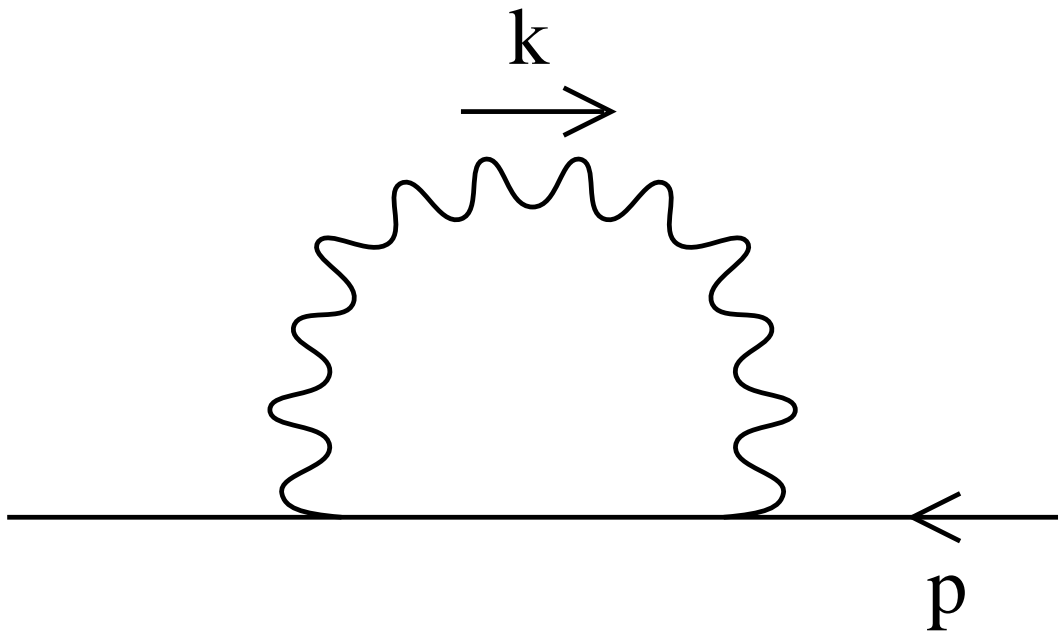
\item[Fig.~1]
One-loop diagram for the quark self-energy. 
$k$ is the loop momentum and $p$ is the external quark momentum.

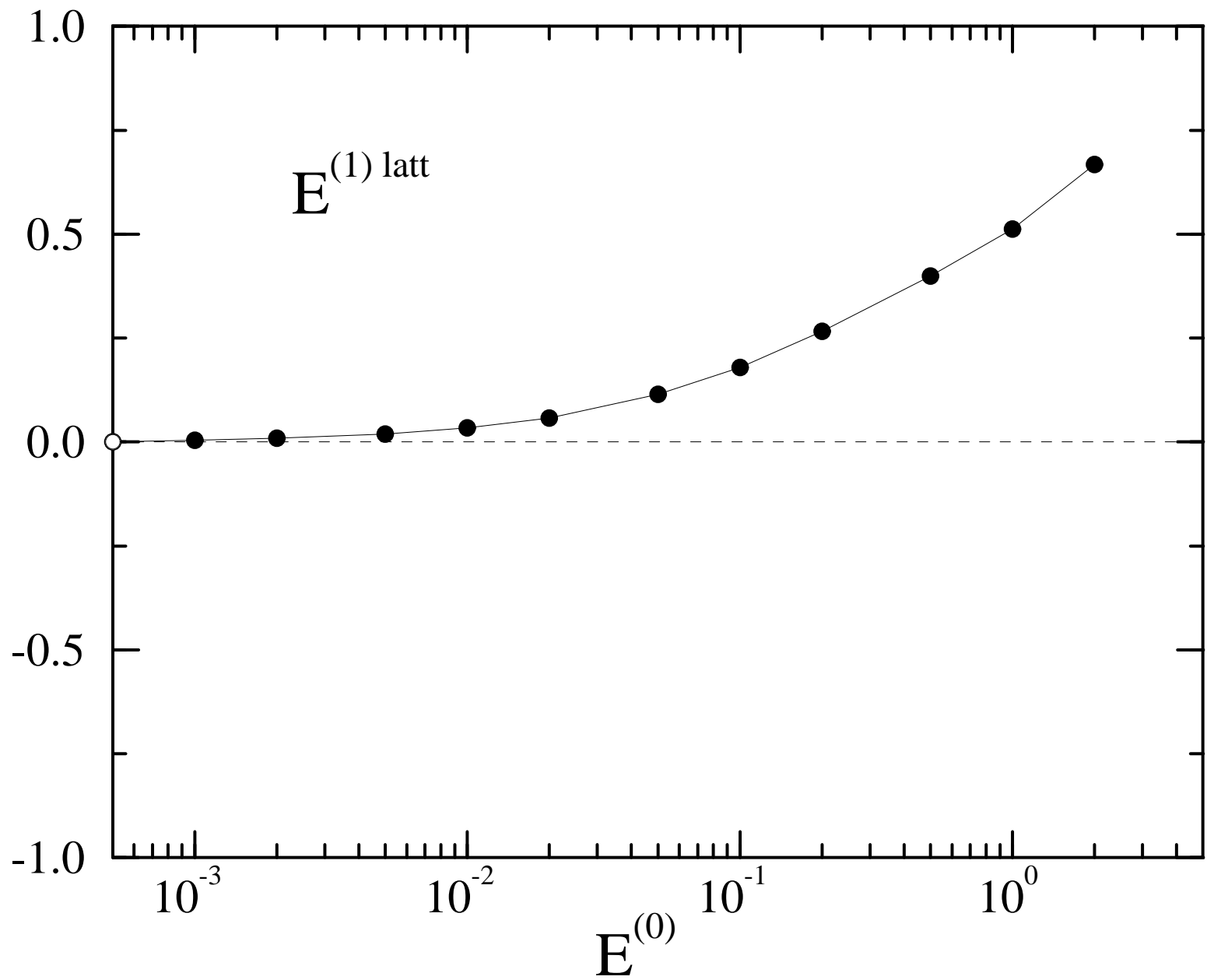
\item[Fig.~2]
One-loop correction to the pole mass on the lattice
as a function of $E^{(0)}$. The Wilson parameter $r$ is chosen 
to be one. Open symbol denotes the value at $E^{(0)}=0$.

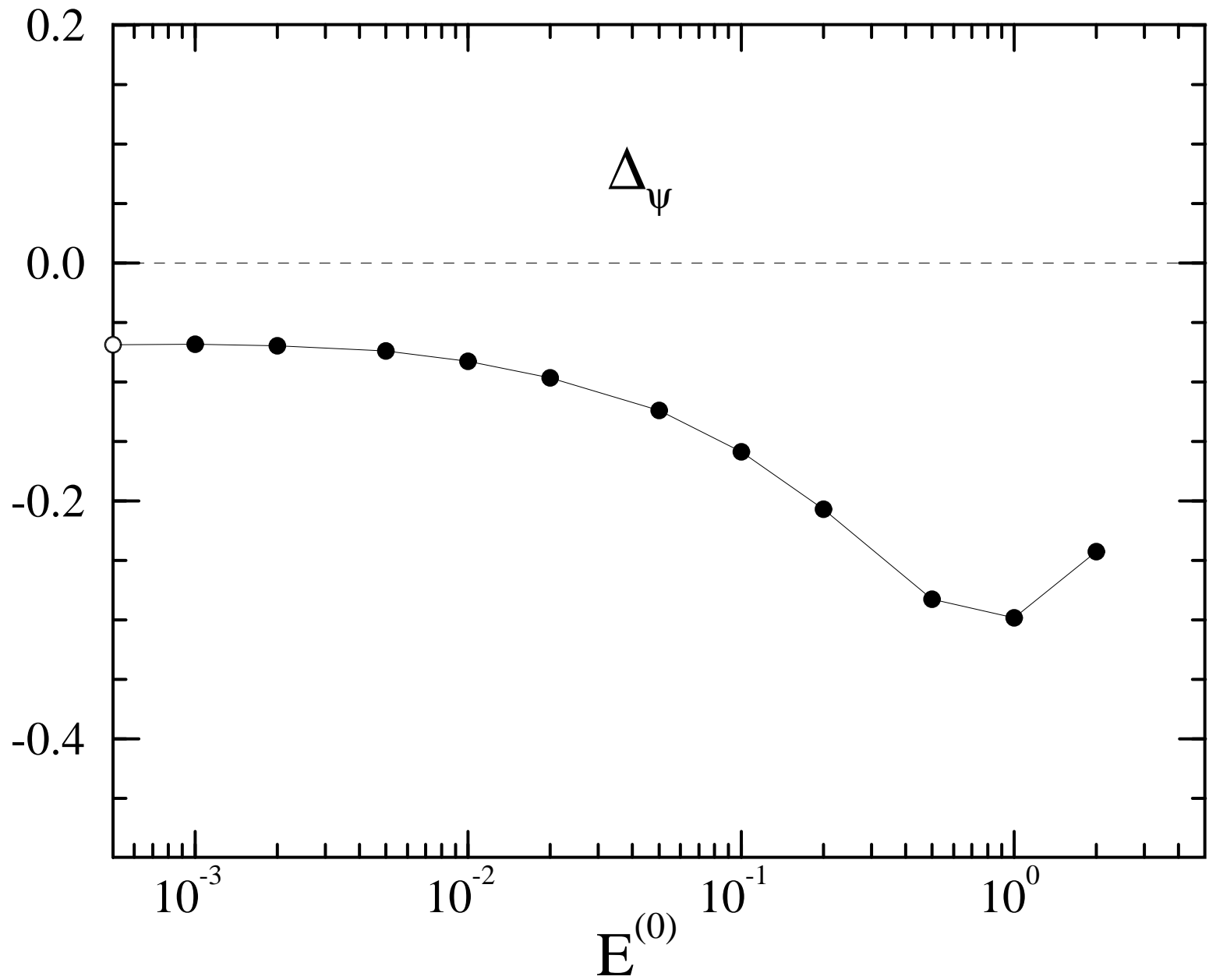
\item[Fig.~3]
One-loop coefficient of the relation 
between the on-shell wave-function renormalization 
factors on the lattice
and in the continuum with NDR scheme
as a function of $E^{(0)}$.
The Wilson parameter $r$ is chosen to be one.
Open symbol denotes the value at $E^{(0)}=0$.

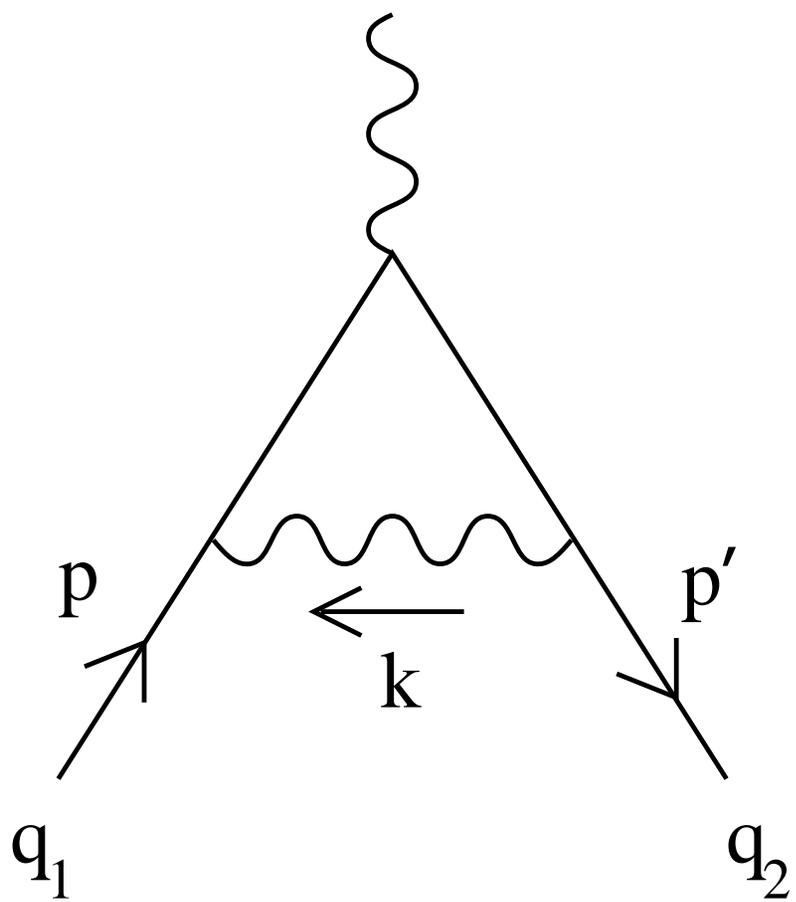
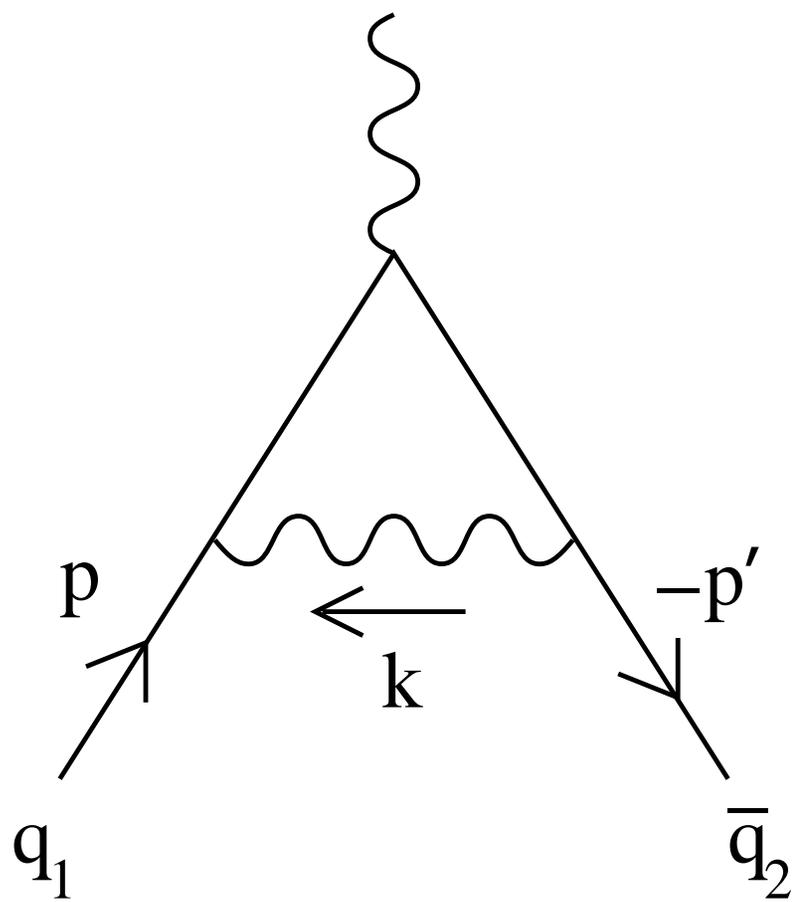
\item[Fig.~4]
One-loop diagrams for the vertex corrections.
$k$ is the loop momentum and $p$ is the incoming 
quark momentum. $p^\prime$ denotes the outgoing quark momentum 
for (a) and the incoming anti-quark momentum for (b).

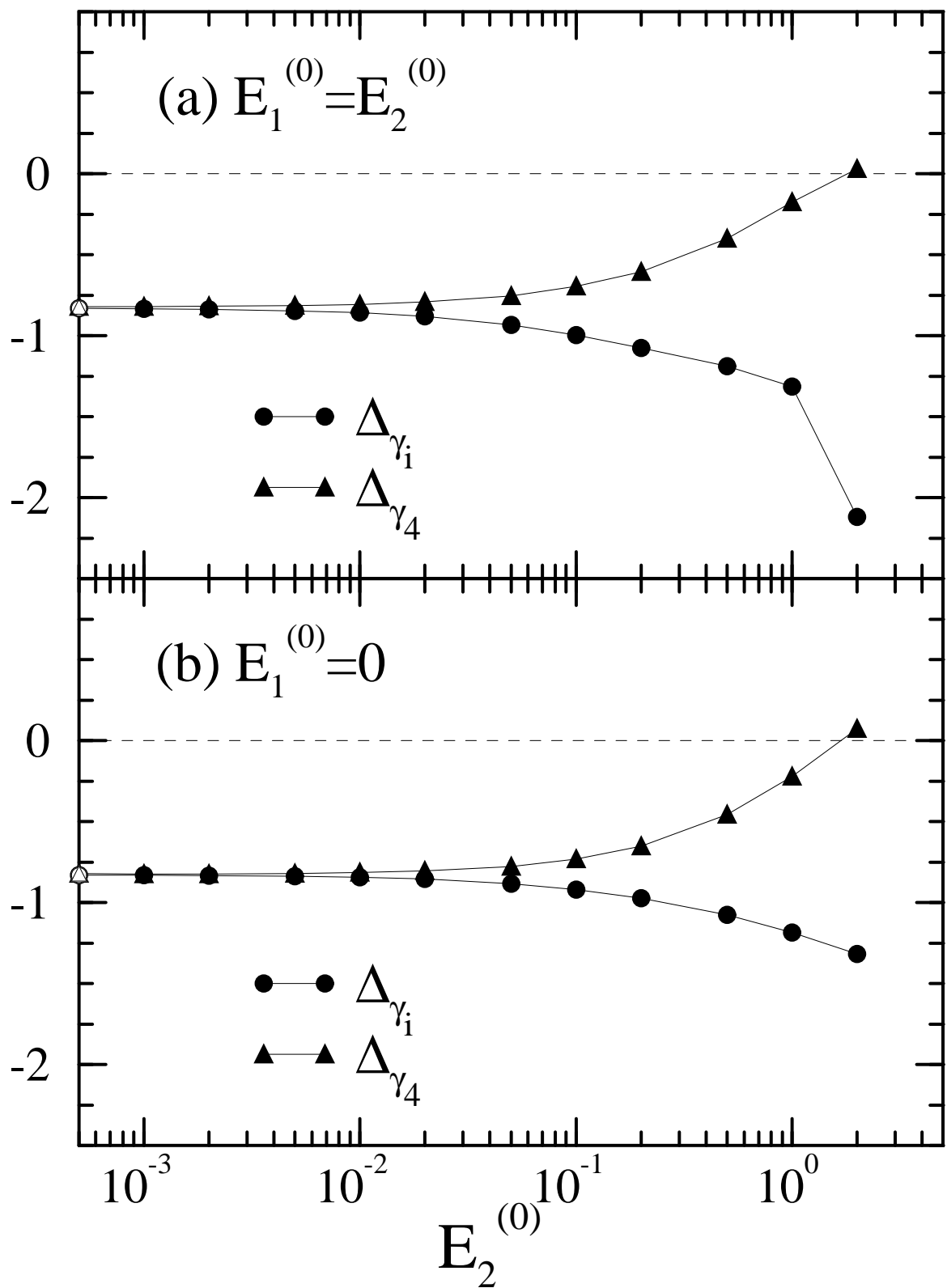
\item[Fig.~5]
One-loop coefficients of the relation 
between the vertex corrections for $\gamma_\mu$ on the lattice
and in the continuum with NDR scheme.
(a) is for the case of $E_1^{(0)}=E_2^{(0)}$ and
(b) for $E_1^{(0)}=0$. 
The Wilson parameter $r$ is chosen to be one.
Open symbols denote the values at $E^{(0)}=0$.

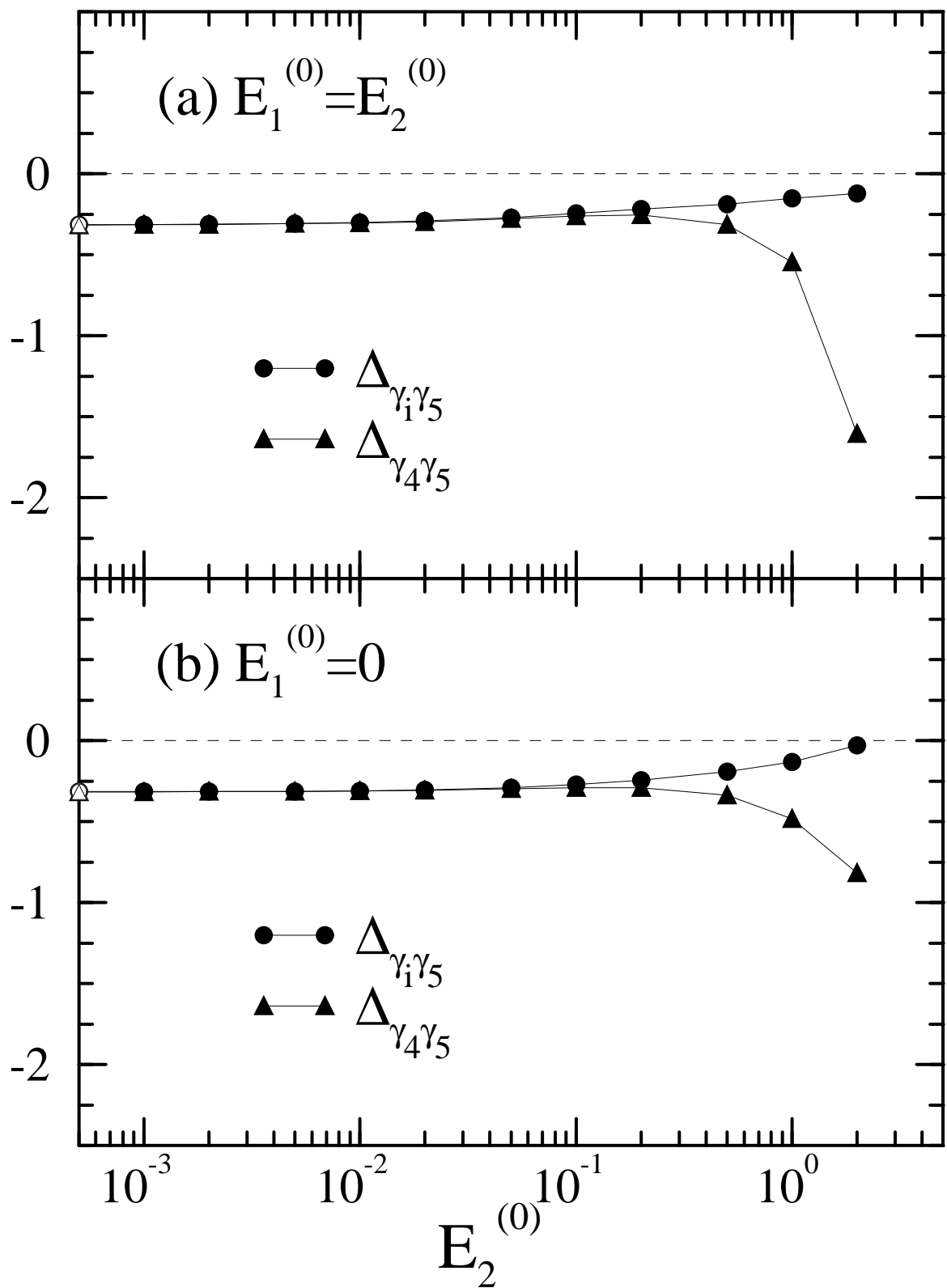
\item[Fig.~6]
One-loop coefficients of the relation 
between the vertex corrections for $\gamma_\mu\gamma_5$ on the lattice
and in the continuum with NDR scheme.
(a) is for the case of $E_1^{(0)}=E_2^{(0)}$ and
(b) for $E_1^{(0)}=0$. 
The Wilson parameter $r$ is chosen to be one.
Open symbols denote the values at $E^{(0)}=0$.

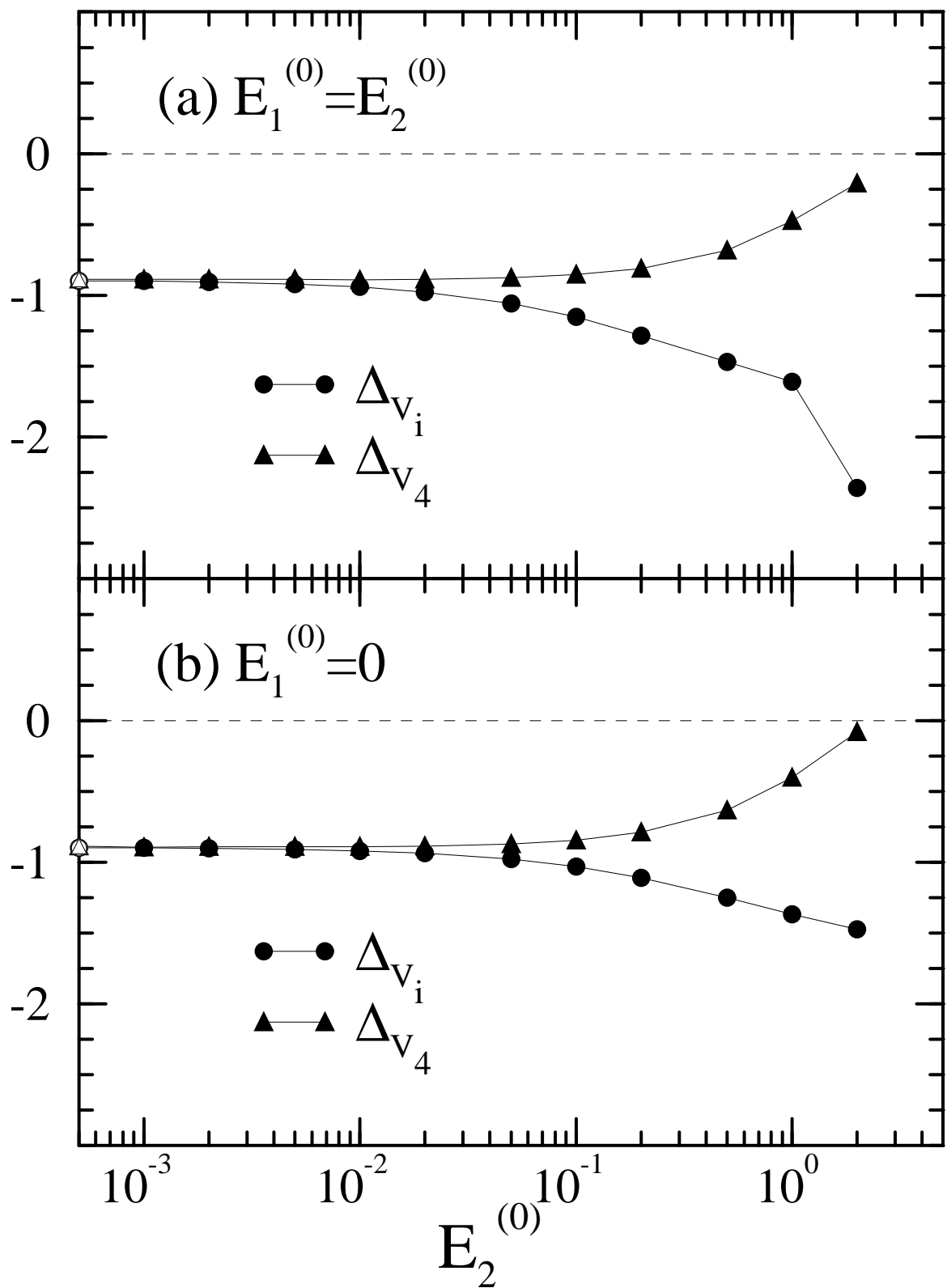
\item[Fig.~7]
One-loop coefficients of the renormalization
factors for vector currents.
(a) is for the case of $E_1^{(0)}=E_2^{(0)}$ and
(b) for $E_1^{(0)}=0$. 
The Wilson parameter $r$ is chosen to be one.
Open symbols denote the values at $E^{(0)}=0$.

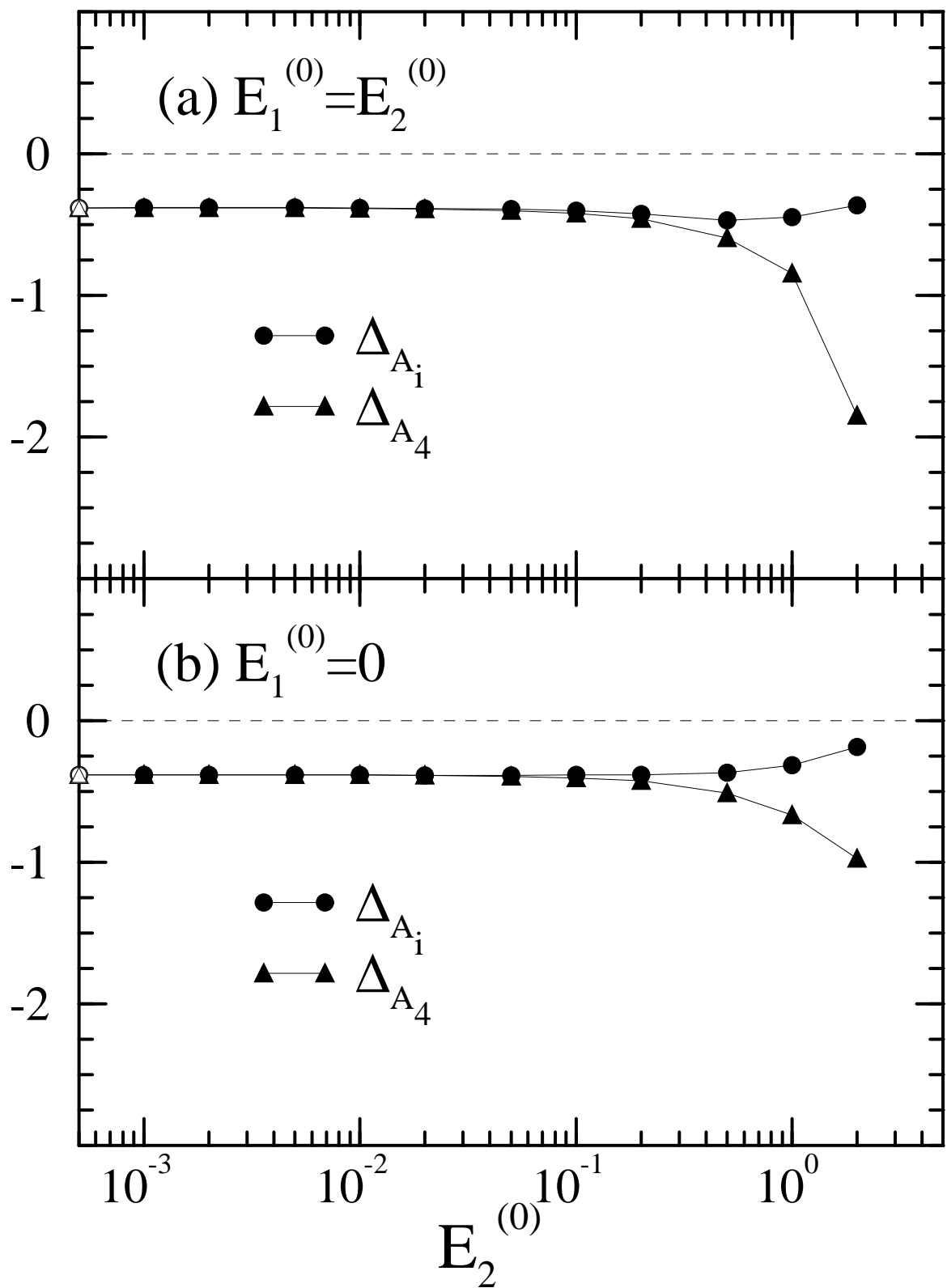
\item[Fig.~8]
One-loop coefficients of the renormalization
factors for axial vector currents.
(a) is for the case of $E_1^{(0)}=E_2^{(0)}$ and
(b) for $E_1^{(0)}=0$. 
The Wilson parameter $r$ is chosen to be one.
Open symbols denote the values at $E^{(0)}=0$.

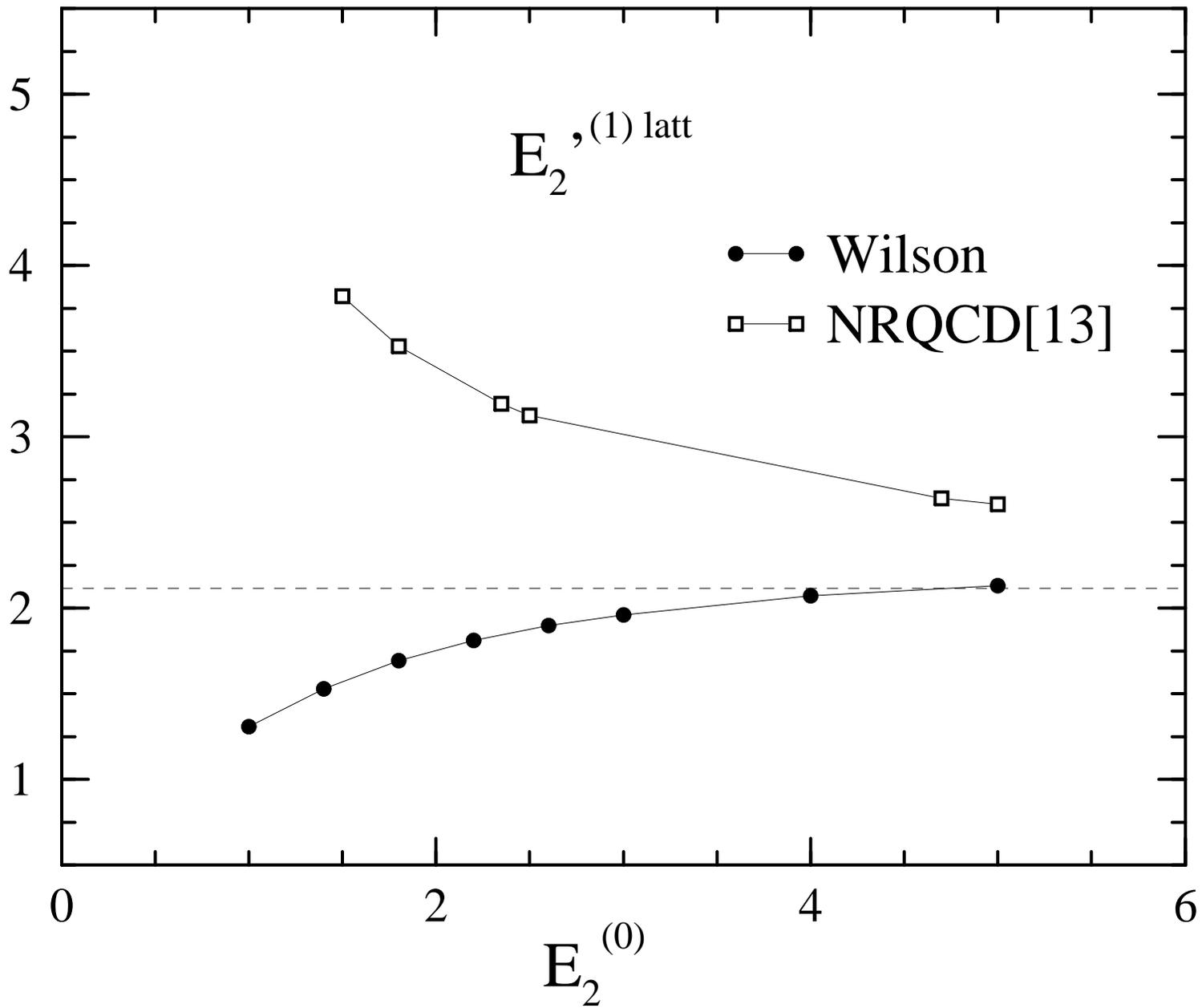
\item[Fig.~9]
$E_2^{(0)}$ dependence of
${{E_2{}^\prime}^{(1)}}^{\ssl}(E_2^{(0)},r)$ for the Wilson
quark action and the NRQCD\protect{\cite{nrqcd1}}. 
The Wilson parameter $r$ is chosen 
to be one. Dotted line denotes the static result in
Ref.~\protect{\cite{static1}}.

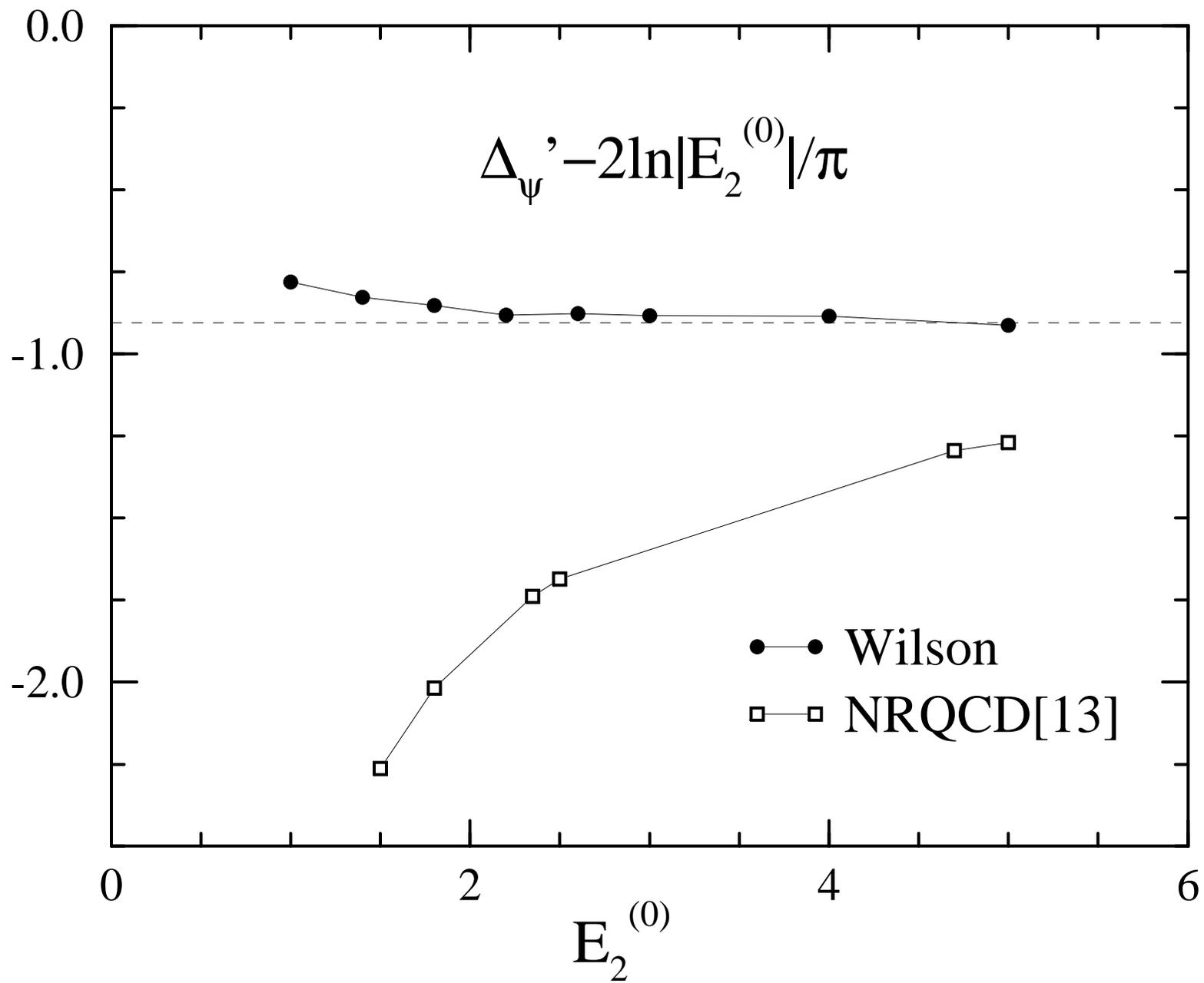
\item[Fig.~10]
$E_2^{(0)}$ dependence of
$\Delta^\prime_\psi(E_2^{(0)},r)-2/\pi\logn{E_2^{(0)}}$
for the Wilson
quark action and the NRQCD\protect{\cite{nrqcd1}}.
The Wilson parameter $r$ is chosen to be one.
Dotted line denotes the static result in
Ref.~\protect{\cite{static1}}.

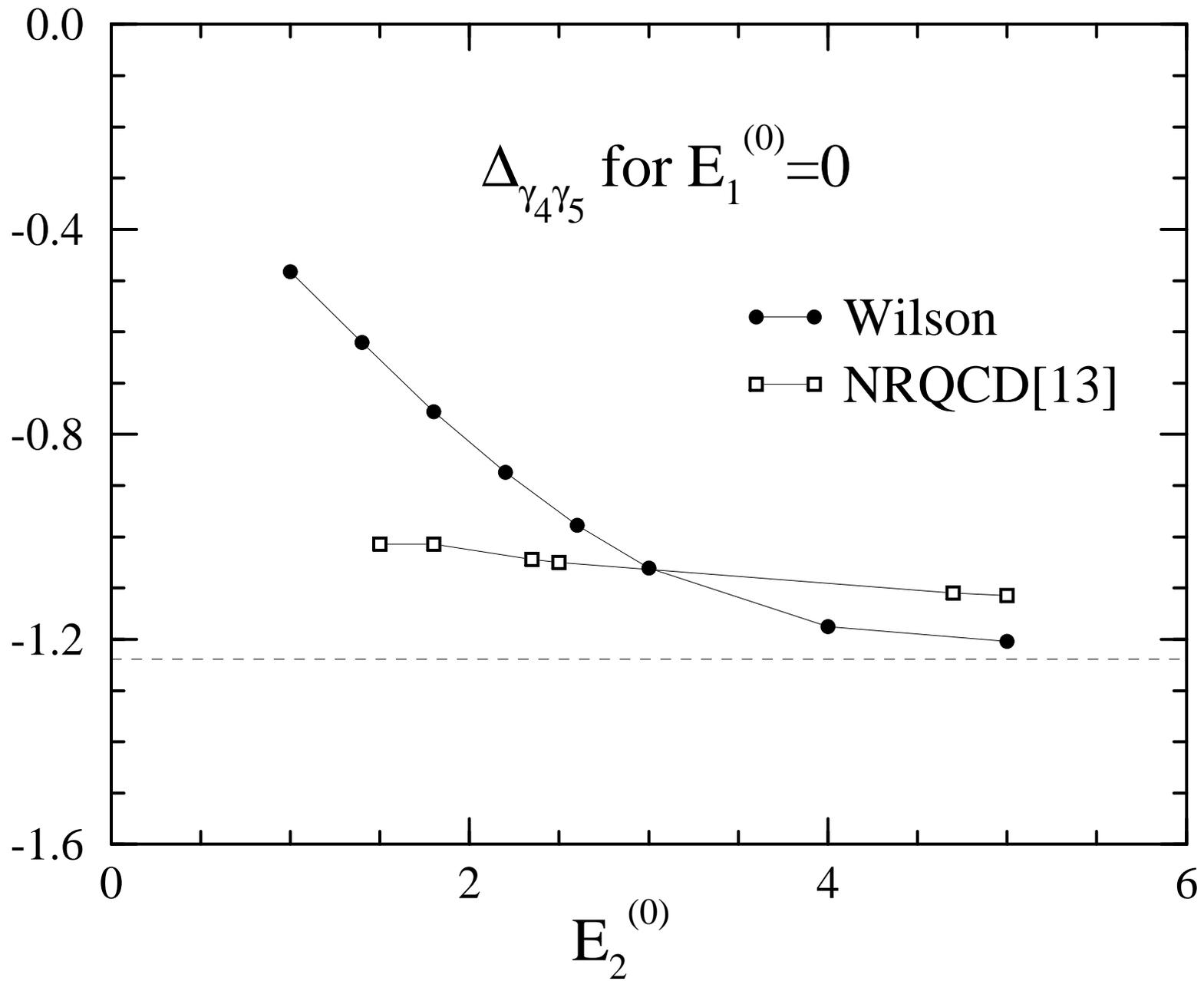
\item[Fig.~11]
$E_2^{(0)}$ dependence of
$\Delta_{\gamma_4\gamma_5}(E_1^{(0)}=0,E_2^{(0)},r)$
for the Wilson
quark action and the NRQCD\protect{\cite{nrqcd1}}.
The Wilson parameter $r$ is chosen to be one.
Dotted line denotes the static result in
Ref.~\protect{\cite{static1}}.

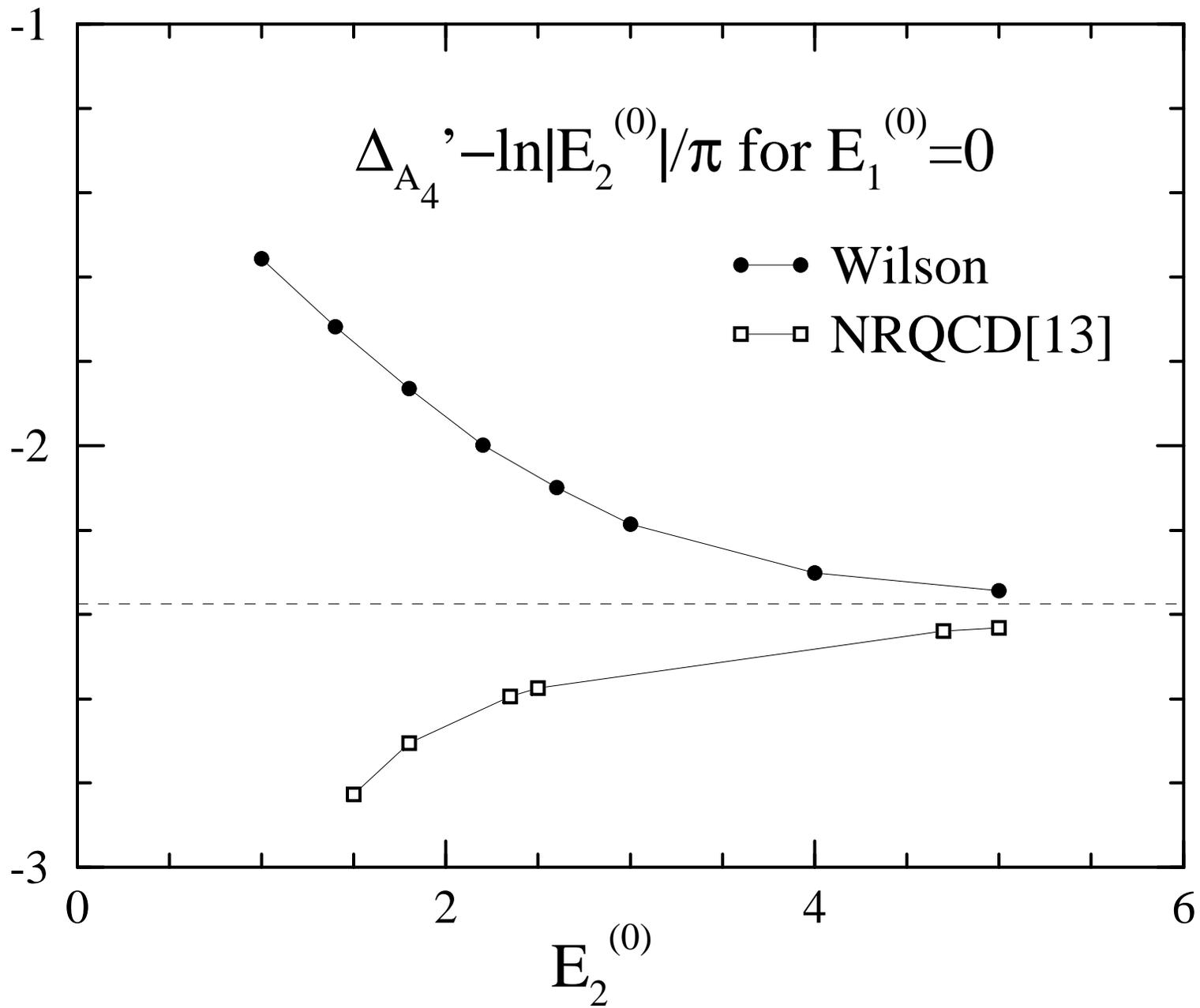
\item[Fig.~12]
$E_2^{(0)}$ dependence of
$\Delta^\prime_{A_4}(E_1^{(0)}=0,E_2^{(0)},r)
-1/\pi\logn{E_2^{(0)}}$
for the Wilson
quark action and the NRQCD\protect{\cite{nrqcd1}}.
The Wilson parameter $r$ is chosen to be one.
Dotted line denotes the static result in
Ref.~\protect{\cite{static1}}.
 
\end{itemize} 
 
\end{document}